\documentclass[a4paper, 11pt, twoside]{article}
\usepackage{CJK}
\usepackage{color}
\usepackage{latexsym,bm}
\usepackage[tbtags]{amsmath}
\usepackage{amssymb}
\usepackage{fancyhdr}
\usepackage{cite}
\usepackage{tabularx}
\usepackage{booktabs}
\usepackage{dcolumn}
\usepackage{graphicx}
\usepackage{wrapfig}
\usepackage{multicol}
\usepackage{verbatim}
\usepackage{mathrsfs}
\usepackage{multirow}

\usepackage{graphicx}

\def\shorttitle{Multi-layer analytic solution for $k\mbox{-}\omega$ model equations}
\def\shortauthor{Fan Tang, Wei-Tao Bi and Zhen-Su She }
\usepackage{Shangda1226}                    

\newcommand{\supercite}[1]{\!\!\textsuperscript{\cite{#1}}} 

\begin{document} 

\begin{CJK*}{GBK}{song}
\title{{\large  \textbf{Multi-layer analytic solution for $k\mbox{-}\omega$ model equations via a symmetry approach}}
\thanks{\footnotesize 
\hspace*{-3.3mm} \dag\,\,Corresponding author, E-mail: she@pku.edu.cn\;
\newline        
Project supported by the NNW Project and the NSF of China under Grant Number 91952201, 11372008, 11452002.
}}
\author{\small{Fan Tang$^{1,2}$,\quad Wei-Tao Bi$^{1,2}$,\quad Zhen-Su She$^{1,2,\dag}$} 
\\[2mm]
\footnotesize{$^1$ Department of Mechanics and Engineering Science, College of Engineering, }
\\\footnotesize{Peking University, Beijing 100871, China}
\\
\footnotesize{$^2$ State Key Laboratory for Turbulence and Complex Systems, Peking University,}
\\\footnotesize{Beijing 100871, China}
\\
\footnotesize{(Received  \,/\, Revised )}\\[2mm]}
\maketitle 
\begin{abstract}
\noindent \textbf{Abstract~~~} Despite being one of the oldest and most widely-used turbulence models in engineering CFD, the $k\mbox{-}\omega$ model has not been fully understood theoretically because of its high nonlinearity and complex model parameter setting. Here, a multi-layer analytic expression is postulated for two lengths (stress and kinetic energy lengths), yielding an analytic solution for the $k\mbox{-}\omega$ model equations in pipe flow. Approximate local balance equations are analyzed to determine key parameters in the solution, which are shown to be rather close to the empirically-measured values from numerical solution of the Wilcox $k\mbox{-}\omega$ model, hence the analytic construction is fully validated. Furthermore, the predictions of three critical locations in the model's three transition functions are validated, which enables an in-depth understanding of the parameter setting of the model. These results provide clear evidence that the $k\mbox{-}\omega$ model sets in it a multi-layer structure, which is similar to but different, in some insignificant details, from the Navier-Stokes turbulence. This finding explains
why the $k\mbox{-}\omega$ model is so popular, especially in computing the near-wall flow. Finally, the analysis is extended to a newly-refined $k\mbox{-}\omega$ model called SED $k\mbox{-}\omega$, showing that the SED $k\mbox{-}\omega$ model has improved the multi-layer structure in the outer flow but preserved the setting of the $k\mbox{-}\omega$ model in the inner region.
\\[2mm]
\textbf{Key words~~~}turbulence model, $k\mbox{-}\omega$ model, structure ensemble dynamics, multi-layer structure, symmetry
\\[2mm]
\end{abstract}


\section{Introduction}\label{sec:introduction}

In computational fluid dynamics (CFD), solving the Reynolds-averaged Navier-Stokes (RANS) equations to predict turbulent flows has been a crucial technique for decades, in which a major bottleneck is turbulence model aiming to construct a closure for the Reynolds stresses\supercite{CFD2030}. Without a universal principle, conventional strategies assume model transport equations for $N$ specific turbulent quantities to model the Reynolds stresses (then called $N$-equation model)\supercite{Wilcox2006}. Numerous coefficients and even functions appear in the $N$ transport equations, to be determined {\it a posteriori} by fitting experimental data. Though being enormously successful for simulating benchmark flows, this mainstream RANS approach encounters a bottleneck for further development, as Spalart pointed out in the 2006 AIAA annual conference\supercite{Spalart2006} with the following characteristics: increasing model complexity for practical engineering flows, insufficient and sometimes undetermined accuracy, etc. For further development, a theoretical understanding of the model equation beyond {\it a posteriori} verification is urgently needed.

A good theoretical understanding of the model equation is often the derivation of analytic solution. This importance is fully recognized by D. C. Wilcox. In his book\supercite{Wilcox2006}, Wilcox stated that
\begin{itemize}
\item[]
Our inability to obtain closed-form solutions is unfortunate because such solutions are invaluable in design studies and for determining trends with a parameter such as Reynolds number, or more generally, for establishing laws of similitude. Furthermore, without analytical solutions, our ability to check the accuracy of numerical solutions is limited.
\end{itemize}
In this work, we conduct an analytic study on the $k\mbox{-}\omega$ model -- one of the oldest and most successful turbulence models in history. In 1942, in addition to $k$ -- the turbulent kinetic energy, Kolmogorov\supercite{Kolmogorov} introduced a second turbulent quantity $\omega$, which he referred to as ``the rate of dissipation of energy in unit volume and time". The reciprocal of $\omega$ is a turbulent time scale, which constitutes a turbulent viscosity $\nu_t$ with $k$ via a dimensional argument: $\nu_t=k/\omega$. Kolmogorov assumed that $\omega$ satisfies a similar transport equation as $k$, which leads to the first two-equation turbulence model known as the $k\mbox{-}\omega$ model. In 1970, without knowing Kolmogorov's work, Saffman\supercite{Saffman} formulated also a $k\mbox{-}\omega$ model, which was later improved by Spalding\supercite{Spalding}. During several decades thereafter, Wilcox and his collaborators continuously pursued the path with many successful developments in applying the model to various realistic flows, which were excellently summarized in his book\supercite{Wilcox2006}. So far, the most widely-used $k\mbox{-}\omega$ model is the Wilcox $k\mbox{-}\omega$ model developed in 2006\supercite{Wilcox2006}. Furthermore, the $k\mbox{-}\omega$ model is the core of Menter's SST model\supercite{MenterSST} -- possibly the most popular model in industry, which employs the $k\mbox{-}\omega$ model to compute the near-wall flow and applies the $k\mbox{-}\epsilon$ model to compute the free shear layer.

Then, a basic question arises, but has not yet been answered in the literature: What is the mathematical structure of the solution to the $k\mbox{-}\omega$ model equations, so that it yields an accurate description of the near-wall mean profile of boundary layer? Being highly nonlinear, the $k\mbox{-}\omega$ model equations can not be solved by conventional analytic tools such as the perturbation method. Solution, however, can be constructed under some assumed principle, and then validated. In this work, we choose the simplest wall flow, a pipe flow, which is strictly one-dimensional for the mean fields, to illustrate the validity of a construction under the concept of dilation symmetry-breaking. In our recently proposed structural ensemble dynamics (SED) theory\supercite{SEDPart1,SEDPart2} for wall turbulence, it is assumed that the constraint by the wall on the turbulent statistics can be expressed by a simple symmetry-breaking form (i.e. ansatz) of dilation group invariants of eddy lengths, which yields excellent descriptions of experimentally and numerically observed mean profiles of the canonical Navier-Stokes (N-S) wall turbulence (i.e. channel, circular pipe, and turbulent boundary layer (TBL))\supercite{SEDPart1,SEDPart2}. Because the $k\mbox{-}\omega$ model equations are most successfully used by engineers to compute the near-wall flows, it is reasonable to assume that the above assumption of generalized dilation invariant lengths is also applicable to the $k\mbox{-}\omega$ model equations. \textcolor{black}{Here, the assumption is studied for the pipe flow produced by the Wilcox $k\mbox{-}\omega$ model, for understanding the mathematical structure of the $k\mbox{-}\omega$ model equations. Note that choosing the pipe flow is an appropriate simplification, because the essential features of the near-wall flows are the significant and extraordinary flow variations in the wall-normal direction rather than those in the streamwise and spanwise directions, which are fully captured by the pipe flow. For more complex flows with strong streamwise variations, such as transitional boundary layers\supercite{SEDSLSC,LiuF2021}, TBLs with strong adverse pressure gradients, and even separated boundary layers (to report elsewhere), the same dilation symmetry breaking principle can also be validated for accurately describing the streamwise flow developments, which could allow more constructions of analytic solutions of the $k\mbox{-}\omega$ model, but beyond the scope of the current study as the first such attempt.}

In this paper, we first analyze the stress and kinetic energy lengths to construct a multi-layer analytic description for the pipe flow. Then, we illustrate a globally-directed local analysis on the $k$ and $\omega$ equations, to determine the parameters in the multi-layer solution, which then forms a complete analytic solution to the Wilcox $k\mbox{-}\omega$ model equations for pipe flow. The solution is validated through comparisons with the numerical results of the Wilcox $k\mbox{-}\omega$ model. These results provide clear evidence that the $k\mbox{-}\omega$ model sets in it a multi-layer structure, which exists in both the N-S equations (as validated in the SED theory) and the $k\mbox{-}\omega$ model equations, being thus a universal structure for wall-bounded turbulent flows. Furthermore, by applying the analytic solution we accurately predict the critical locations in the transition functions of the Wilcox $k\mbox{-}\omega$ model, which enables an in-depth understanding of the parameter setting of the model. Finally, the analysis is extended to a newly-refined $k\mbox{-}\omega$ model: the SED $k\mbox{-}\omega$ model proposed by Chen et al.\supercite{SED-k-omega}, showing that the SED $k\mbox{-}\omega$ model has improved the multi-layer structure in the outer flow but preserved the setting of the $k\mbox{-}\omega$ model in the inner region.

The paper is organized as follows. Sect. \ref{sec:k-omega eq} introduces the Wilcox $k\mbox{-}\omega$ model for pipe flow. Sect. \ref{sec:analytic solution} describes, respectively, the construction of the analytic solution, the theoretical determination of the parameters in the solution, the predictions about the transition functions in the Wilcox $k\mbox{-}\omega$ model, as well as the numerical validation. Sect. \ref{sec:improved model} extends the analysis to the SED $k\mbox{-}\omega$ model. Finally, Sect. \ref{sec:conclusion} discusses and concludes the study.

\section{$k\mbox{-}\omega$ model equations}\label{sec:k-omega eq}
The $k$ and $\omega$ equations of Wilcox (2006)\supercite{Wilcox2006} read:
\begin{align}
\label{eq:k3d} &\frac{\partial k}{\partial t}+U_j\frac{\partial k}{\partial x_j}=\tau_{ij}\frac{\partial U_i}{x_j}-\beta^*k\omega+\frac{\partial}{\partial x_j}\left[\left(\nu+\sigma^*\nu_t\right)\frac{\partial k}{\partial x_j}\right],\\[1mm]
\label{eq:omega3d} &\frac{\partial \omega}{\partial t}+U_j\frac{\partial \omega}{\partial x_j}=\alpha\frac{\omega}{k}\tau_{ij}\frac{\partial U_i}{x_j}-\beta\omega^2+\frac{\partial}{\partial x_j}\left[\left(\nu+\sigma\nu_t\right)\frac{\partial \omega}{\partial x_j}\right]+\frac{\sigma_d}{\omega}\frac{\partial k}{\partial x_j}\frac{\partial \omega}{\partial x_j}.
\end{align}
In (\ref{eq:k3d}) and (\ref{eq:omega3d}), the right hand side consists of production, dissipation, viscous diffusion and turbulent transport in order, and there is an additional cross-diffusion term in (\ref{eq:omega3d}). \textcolor{black}{In the equations, $U_j$ is the $j$-th component of mean velocity ($j=1,2,3$), $x_j$ is the Cartesian coordinate, $t$ is time, $\tau_{ij}$ is the Reynolds stress, $\nu$ is the molecular kinematic viscosity, and $\nu_t$ is the eddy viscosity. The model parameters (e.g. $\beta^*$, $\sigma^*$, $\alpha$) are to be explained in below.} Note that $k$ denotes the total turbulent kinetic energy, i.e. the sum of the streamwise, wall-normal and spanwise kinetic energy components. However, as addressed by Wilcox\supercite{Wilcox2006}, it is not critically important whether $k$ is taken to be the full kinetic energy, or, alternatively, the streamwise component only. In what follows, we assume that $k={\left \langle uu \right \rangle}/2$ where $u$ is the streamwise velocity fluctuation and ${\left \langle \right \rangle}$ denotes Reynolds average.

Here, we consider the fully-developed turbulent pipe flow, such that the above two equations in the wall-normal coordinate reduce to\supercite{SED-k-omega}:
\begin{align}
\label{eq:k1d} &S^+W^+-\beta^*k^+\omega^++\frac{1}{r}\frac{d }{dy^+}\left[r\left(1+\sigma^*\nu_t^+\right)\frac{dk^+}{dy^+}\right]=0,\\[1mm]
\label{eq:omega1d} &\alpha S^+W^+-\beta k^+\omega^++\frac{1}{r}\frac{\nu_t^+}{\alpha^*}\frac{d }{dy^+}\left[r\left(1+\sigma\nu_t^+\right)\frac{d\omega^+}{dy^+}\right]+\sigma_d \frac{k^+}{\omega^{+2}}\frac{dk^+}{dy^+}\frac{d\omega^+}{dy^+}=0,
\end{align}
where $y^+$ is the dimensionless wall-normal coordinate, $r=1-y/\delta=1-y^+/Re_\tau$ ($\delta$ is the pipe radius and $Re_\tau$ the friction Reynolds number) is the dimensionless distance away from the pipe center, superscript plus denotes wall-unit normalization, $S^+=dU^+/dy^+$ is the mean shear ($U$ is the streamwise mean velocity), $W^+=-{\left \langle uv \right \rangle}^+$ is the Reynolds shear stress ($v$ is the wall-normal component of the fluctuating velocity). The eddy viscosity $\nu_t^+$ reads:
\begin{align}
\label{eq:nut} &\nu_t^+=\frac{W^+}{S^+}=\alpha^*\frac{k^+}{\omega^+}\equiv\alpha^*R^+,
\end{align}
where $R^+\equiv k^+/\omega^+$ is a quantity which the $k\mbox{-}\omega$ model uses to introduce transition (from inner flow to outer flow) so as to formulate a multi-layer structure, as we see below. The wall-normally integrated streamwise mean momentum equation reads\supercite{SED-k-omega}:
\begin{align}
\label{eq:momentum} &S^++W^+=r.
\end{align}
The model parameters are set as follows. In (\ref{eq:k1d}) and (\ref{eq:omega1d}), $\sigma^*=0.6$, $\sigma=0.5$, $\beta=0.0708$, and $\sigma_d=0.125$. $\beta^*$, $\alpha^*$ and $\alpha\alpha^*$ are three transition functions defined as:
\begin{align}
\label{eq:Rbeta} &\beta^*=\beta_0^*\frac{100\beta_0/27+(R^+/R_\beta)^4}{1+(R^+/R_\beta)^4},\\[1mm]
\label{eq:Rk} &\alpha^*=\frac{\alpha_0^*+R^+/R_k}{1+R^+/R_k},\\
\label{eq:Romega} &\alpha\alpha^*=\alpha_\infty\frac{\alpha_0+R^+/R_\omega}{1+R^+/R_\omega},
\end{align}
where $\beta_0^*=0.09$, $\beta_0=\beta=0.0708$, $\alpha_0^*=\beta_0/3=0.0236$, $\alpha_\infty=0.52$, $\alpha_0=1/9$. The most interesting quantities are three transition thresholds: $R_\beta=8$, $R_k=6$, and $R_\omega=2.61$.

Wilcox has been fully aware that a multi-layer structure is implicitly set in the $k\mbox{-}\omega$ model. In his book\supercite{Wilcox2006} a local perturbation analysis on the $k\mbox{-}\omega$ model equations (and on other turbulence models) has been performed for the viscous sublayer, log-layer, and defect layer, respectively. In the log-layer, specifically, a local analytic solution has been derived by Wilcox for the $k\mbox{-}\omega$ model equations, which reads:
\begin{align}
\label{eq:k_log} &k^+\approx1/\sqrt{\beta_0^*},\\[1mm]
\label{eq:omega_log} &\omega^+\approx\frac{1}{\sqrt{\beta_0^*}\kappa y^+},
\end{align}
which yields the celebrated log-law of the wall:
\begin{align}
\label{eq:U_log} &U^+=\frac{1}{\kappa}\text{ln} y^++C,
\end{align}
where
\begin{align}
\label{eq:kappa} &\kappa=\sqrt{\frac{\beta_0-\alpha_\infty\beta_0^*}{\sigma\sqrt{\beta_0^*}}}.
\end{align}
From the parameter values by Wilcox cited above, one obtains $\kappa=0.4$.

\section{A multi-layer analytic solution to the $k\mbox{-}\omega$ model equations}\label{sec:analytic solution}
A fundamental question in the theoretical study of wall turbulence is how to quantify the mean flow property as a function of flow condition, and a promising road is to study it through the concept of invariance or similarity. In the canonical wall-bounded turbulent flows, owing to the solid wall, only the dilation invariance group exists in the non-trivial wall-normal direction. As explained in \cite{SEDPart1,SEDPart2}, when the dilation group invariance encounters a symmetry-breaking because of variation of balance mechanisms for turbulent fluctuations, a generalized dilation invariance principle can be formulated to yield, in the wall-normal direction, a continuous four-layer structure describing the viscous sublayer, buffer layer, log-layer, and wake region, in order. This generalized dilation invariance reads: $F(x)=cx^\alpha\left[1+\left(x/x_c\right)^p \right]^{\gamma/p}=cx^\alpha \Pi(x)$, which formulates a power-law jump from $x^\alpha$ to $x^{\alpha+\gamma}$ at $x_c$, with $p$ called transition sharpness, often taken to be a big positive integer like 4. Here $\Pi(x)=\left[1+\left(x/x_c\right)^p\right]^{\gamma/p}$ is called a universal dilation ansatz, or dilation-symmetry-breaking principle because it is quite universal.

Note that choosing the stress length and kinetic energy length (instead of the mean velocity) is crucial to identify the similarity structure of wall turbulence. These lengths are argued\supercite{SEDPart1,SEDPart2} to be the right variables to display the simplest dilation-symmetry-breaking to form a multi-layer structure, which, as is shown below, can be expressed with a product of multiple factors, each for one layer, to describe the whole profile. This is a signature of the self-organization principle for the ensemble of turbulent eddies in TBL under the constraint of the wall, so that the flow properties in different layers are similarly linked.

We will see below that the generalized dilation invariance principle continues to hold when one formulates the turbulent pipe flows produced by the Wilcox $k\mbox{-}\omega$ model equations. As a consequence, an analytic solution can be constructed for the Wilcox $k\mbox{-}\omega$ model equations in the pipe flow, which, validated with numerical simulations, reveals a striking fact that the multi-layer structure is a universal structure in the wall-bounded turbulent flows predicted by both the N-S equations and the $k\mbox{-}\omega$ model equations.

\subsection{Constructing multi-layer length functions to the $k\mbox{-}\omega$ model equations}\label{subsec:analytic solution}
In the SED theory the stress length and kinetic energy length are defined, respectively, as:
\begin{align}
\label{eq:ell12} &\ell_{12}^+=\sqrt{W^+}/S^+,\\[1mm]
\label{eq:ellk} &\ell_k^+=\sqrt{2k^+}/S^+.
\end{align}
Once $\ell_{12}^+$ and $\ell_k^+$ are formulated, $S^+$, $W^+$ and $k^+$ can be calculated from Eqs. (\ref{eq:momentum}), (\ref{eq:ell12}) and (\ref{eq:ellk}) to give:
\begin{align}
\label{eq:Splus} &S^+=\frac{-1+\sqrt{1+4r{\ell_{12}^+}^2}}{2{\ell_{12}^+}^2},\\[1mm]
\label{eq:Wplus} &W^+={\ell_{12}^+}^2{S^+}^2,\\
\label{eq:kplus} &k^+=\frac{1}{2}{\ell_k^+}^2{S^+}^2.
\end{align}
Applying the above expressions to the constitutive equation (\ref{eq:nut}) yields:
\begin{align}
\label{eq:nut+} &\nu_t^+={\ell_{12}^+}^2S^+,\\[1mm]
\label{eq:omega+} &\omega^+=\frac{1}{2}\alpha^*S^+\left(\frac{\ell_k^+}{\ell_{12}^+}\right)^2.
\end{align}
Note that Eqs. (\ref{eq:Splus})-(\ref{eq:omega+}) are exact in the whole boundary layer, and (\ref{eq:omega+}) is an implicit expression.

The SED theory has proposed that both $\ell_{12}^+$ and $\ell_k^+$ possess multi-layer dilation invariance with $y^+$, which can be written as the following for the $k\mbox{-}\omega$ model equations:
\begin{align}
\label{eq:ell12formula} &\ell_{12}^+=\ell_0^+\left(\frac{y^+}{y_{sub}^+}\right)^2\left[1+\left(\frac{y^+}{y_{sub}^+}\right)^8\right]^{\gamma_w/8}
\left[1+\left(\frac{y^+}{y_{buf}^+}\right)^2\right]^{-(1+\gamma_w)/2}
\frac{1-r^m}{m(1-r)}\frac{1}{Z_c}\left[1+\left(\frac{r_c}{r}\right)^2\right]^{1/4},\\[1mm]
\label{eq:ellkformula} &\ell_k^+=\ell_{k0}^+\left(\frac{y^+}{y_{ksub}^+}\right)\left[1+\left(\frac{y^+}{y_{ksub}^+}\right)^4\right]^{\gamma_k/4}
\left[1+\left(\frac{y^+}{y_{kbuf}^+}\right)^4\right]^{-\gamma_k/4}
\frac{1-r^m}{m(1-r)}\frac{1}{Z_{kc}}\left[1+\left(\frac{r_{kc}}{r}\right)^2\right]^{1/2},
\end{align}
where
\begin{align}
\label{eq:ell0+} &\ell_0^+=\frac{\kappa{y_{sub}^+}^{2+\gamma_w}}{{y_{buf}^+}^{1+\gamma_w}},\qquad
\ell_{k0}^+=\frac{\kappa_k{y_{ksub}^+}^{1+\gamma_k}}{{y_{kbuf}^+}^{\gamma_k}},\\[1mm]
\label{eq:Zwk} &Z_c=\left(1+{r_c}^2\right)^{1/4},\qquad Z_{kc}=\left(1+{r_{kc}}^2\right)^{1/2}.
\end{align}
There are several parameters in (\ref{eq:ell12formula}) and (\ref{eq:ellkformula}), which are called multi-layer structure parameters, to be determined theoretically and empirically for the $k\mbox{-}\omega$ model. In (\ref{eq:ell12formula}) $y_{sub}^+$ is the sublayer thickness for $\ell_{12}^+$, beneath which $\ell_{12}^+\propto {y^+}^2$, set by the $k\mbox{-}\omega$ model equations; $y_{buf}^+$ is the buffer layer thicknesses for $\ell_{12}^+$ and in between $y_{sub}^+$ and $y_{buf}^+$, $\ell_{12}^+\propto {y^+}^{2+\gamma_k}$ with $\gamma_k$ being the scaling exponent increment; above $y_{buf}^+$ is the log-layer where $\ell_{12}^+=\kappa y^+$, in accordance with Eq. (\ref{eq:U_log}); further away from the log-layer is the defect layer described with the defect law of $1-r^m$, where $m$ is a scaling exponent; at the pipe center $\ell_{12}^+$ diverges, described here with a center core of radius $r_c$ following the SED theory. The formulation of $\ell_k^+$ in Eq. (\ref{eq:ellkformula}) is similar: $\kappa_k$ is a Karman-like constant for $k$; $y_{ksub}^+$ and $y_{kbuf}^+$ are the sublayer and buffer layer thicknesses for $\ell_k^+$, with $\gamma_k$ being the scaling exponent increment in the buffer layer of $\ell_k^+$; and $r_{kc}$ is the center core radius for $\ell_k^+$.

Note that in true turbulent pipe flows described with the N-S equations, the multi-layer structures of $\ell_{12}^+$ and $\ell_k^+$ are more similar than those in (\ref{eq:ell12formula}) and (\ref{eq:ellkformula}), e.g. $y_{sub}^+=y_{ksub}^+$ and $y_{buf}^+=y_{kbuf}^+$ in the N-S turbulence. One refers to \cite{SEDPart1,SEDPart2} for the values of the multi-layer structure parameters determined by the SED theory (e.g. $\kappa=0.45$ in the SED theory) for the canonical wall-bounded turbulent flows. In the current pipe flows predicted by the Wilcox $k\mbox{-}\omega$ model, the multi-layer structure parameters are influenced by the model parameters, which can be theoretically studied by conducting a globally-directed local analysis on the balance equations and length functions, as shown next.

\begin{figure}[h]
\centering \mbox{ \subfigure[]{\includegraphics[width=75mm]{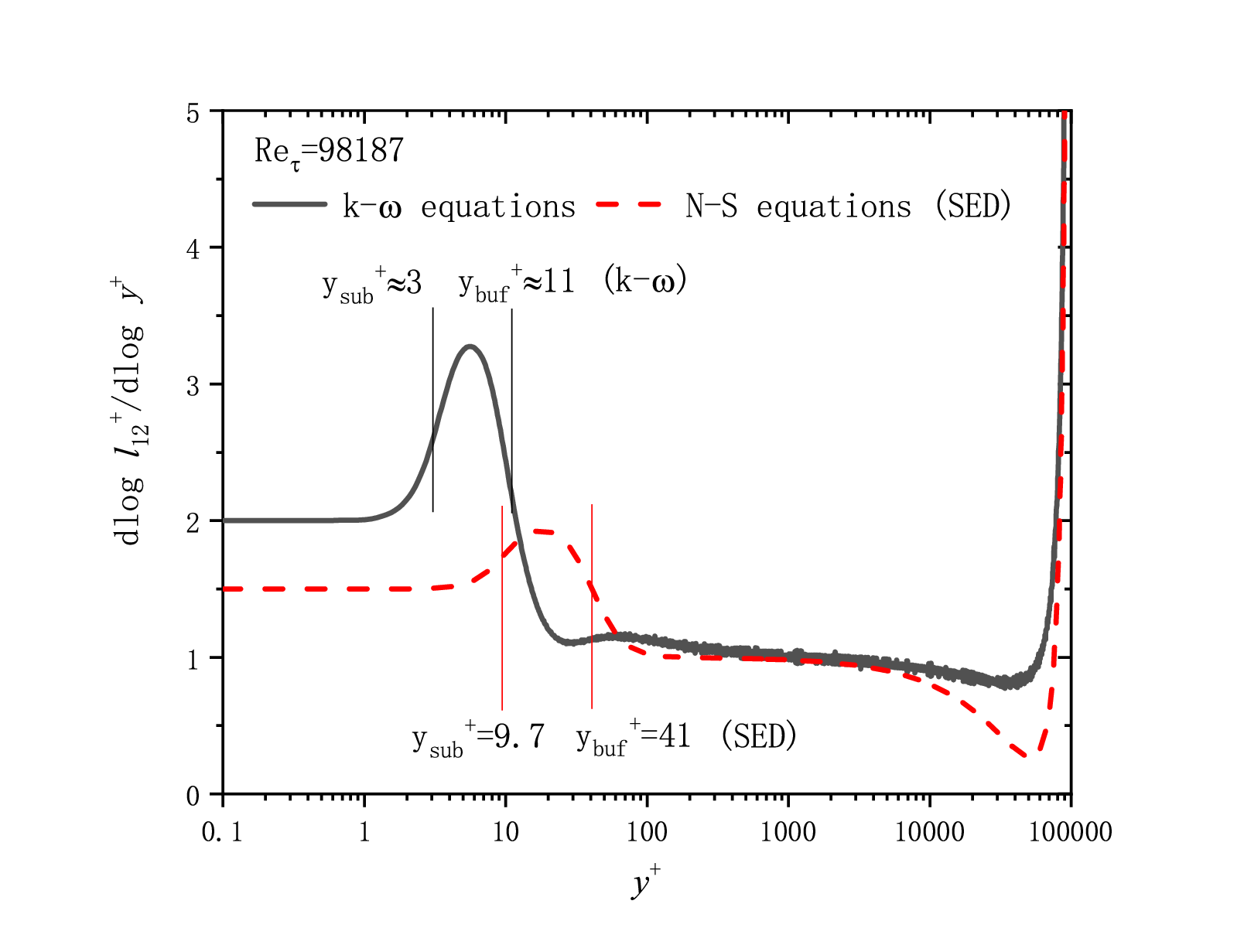}}\quad
\subfigure[]{\includegraphics[width=75mm]{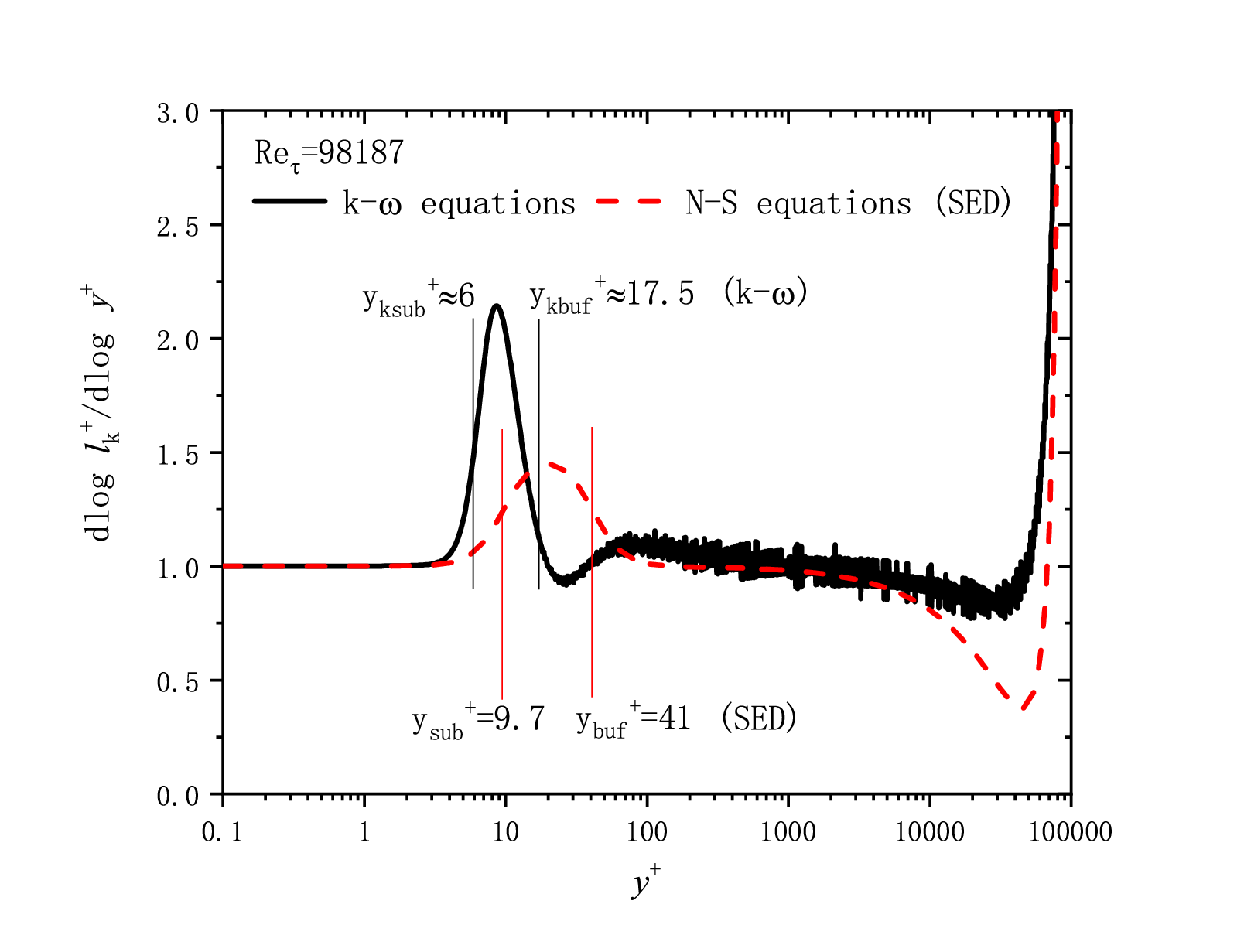}}}
\figcaption{Profiles of the diagnostic functions calculated by the $k\mbox{-}\omega$ model equations for pipe flow at $Re_\tau=98187$, comparing with the corresponding theoretical profiles of the SED. (a) $d\log\ell_{12}^+/d\log y^+$, (b) $d\log\ell_{k}^+/d\log y^+$. The vertical lines indicates the empirical locations of $y_{sub}^+$, $y_{buf}^+$, $y_{ksub}^+$, and $y_{kbuf}^+$ for the $k\mbox{-}\omega$ model, and the theoretical values of $y_{sub}^+$ and $y_{buf}^+$ for the SED.}\label{fig:dlogdlog}
\end{figure}

Before performing the local analysis, let us display the multi-layer structures of the $k\mbox{-}\omega$-predicted pipe flow through plotting the so-called diagnostic functions: $d\log\ell_{12}^+/d\log y^+$ and $d\log\ell_{k}^+/d\log y^+$, which visualize the power laws and the transition of power laws for $\ell_{12}^+$ and $\ell_k^+$ in the multi-layer structures. Note that the data of the $k\mbox{-}\omega$ equations are calculated by employing the companion software provided in Wilcox's book\supercite{Wilcox2006}, and the SED profiles of the diagnostic functions are calculated from the theoretical formula in \cite{SEDPart1,SEDPart2}. As shown in Fig. \ref{fig:dlogdlog}, there are multi-layer structures in both the $k\mbox{-}\omega$-predicted and SED-predicted pipe flows, which consist of, from the left to the right, a flat denoting the viscous sublayer, a plateau denoting the buffer layer, a lower flat denoting the log-layer, a pit denoting the bulk flow, and a divergent center core. However, significant differences occur between the multi-layer structures of the two flows. In the viscous sublayer, the power-law exponent of $\ell_{12}^+$ is 2 for the $k\mbox{-}\omega$ equations, larger than the SED value of 1.5. In the buffer layer, the power-law exponents of both $\ell_{12}^+$ and $\ell_k^+$ are much larger than those of the SED theory. One can estimate that, $\gamma_w\approx 1.5$ and $\gamma_k\approx 1$ (which are taken in Eqs. (\ref{eq:ell12formula}) and (\ref{eq:ellkformula}) hereinafter) for the $k\mbox{-}\omega$ model equations, in contrast to 0.5 in the SED theory for both $\ell_{12}^+$ and $\ell_k^+$. In addition, for the $k\mbox{-}\omega$ equations, $y_{sub}^+$ and $y_{ksub}^+$, as well as $y_{buf}^+$ and $y_{kbuf}^+$, are different from each other, and significantly smaller than the corresponding SED values, as shown with the vertical lines in Fig. \ref{fig:dlogdlog}. Finally, for the SED-predicted pipe flow, there is a pit in the bulk flow regime for both $d\log\ell_{12}^+/d\log y^+$ and $d\log\ell_{k}^+/d\log y^+$, whereas in the $k\mbox{-}\omega$-predicted flow, the pit is rather mild, if exits. Despite these differences, Eqs. (\ref{eq:ell12formula}) and (\ref{eq:ellkformula}) can describe the wall-normally multi-layer structures of the $k\mbox{-}\omega$-predicted pipe flow, as shown below.

\subsection{A SED-directed local analysis to determine the multi-layer parameters}\label{subsec:local analysis}

Now we present a theoretical analysis to determine the multi-layer structure parameters in Eqs. (\ref{eq:ell12formula}) and (\ref{eq:ellkformula}), which is conducted by considering the local approximations of both the balance equations and length functions.

\noindent\textbf{A. Determining $y_{buf}^+$ from a log-layer approximation}

\begin{figure}[h]
\centering \mbox{ \subfigure[]{\includegraphics[width=75mm]{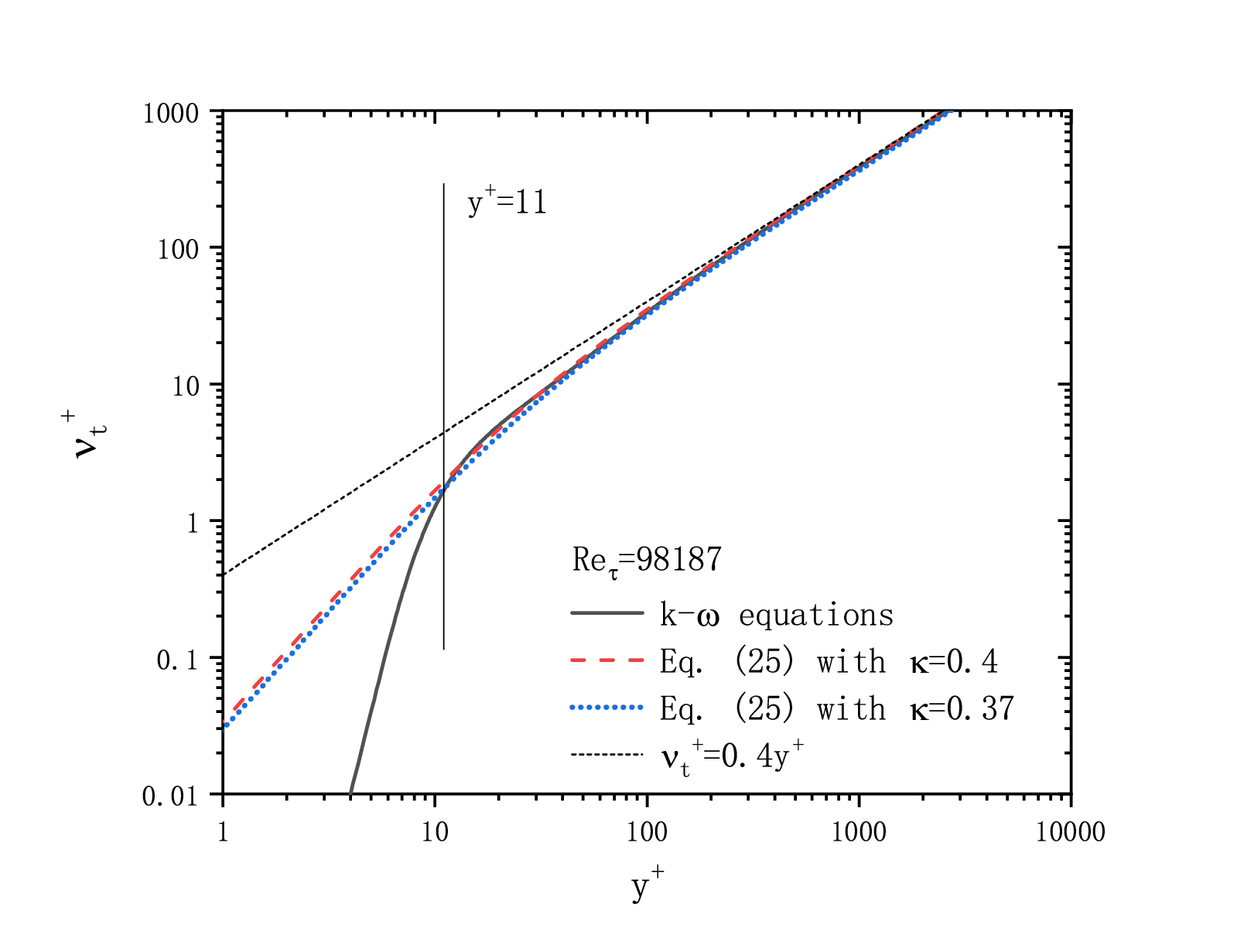}}\quad
\subfigure[]{\includegraphics[width=75mm]{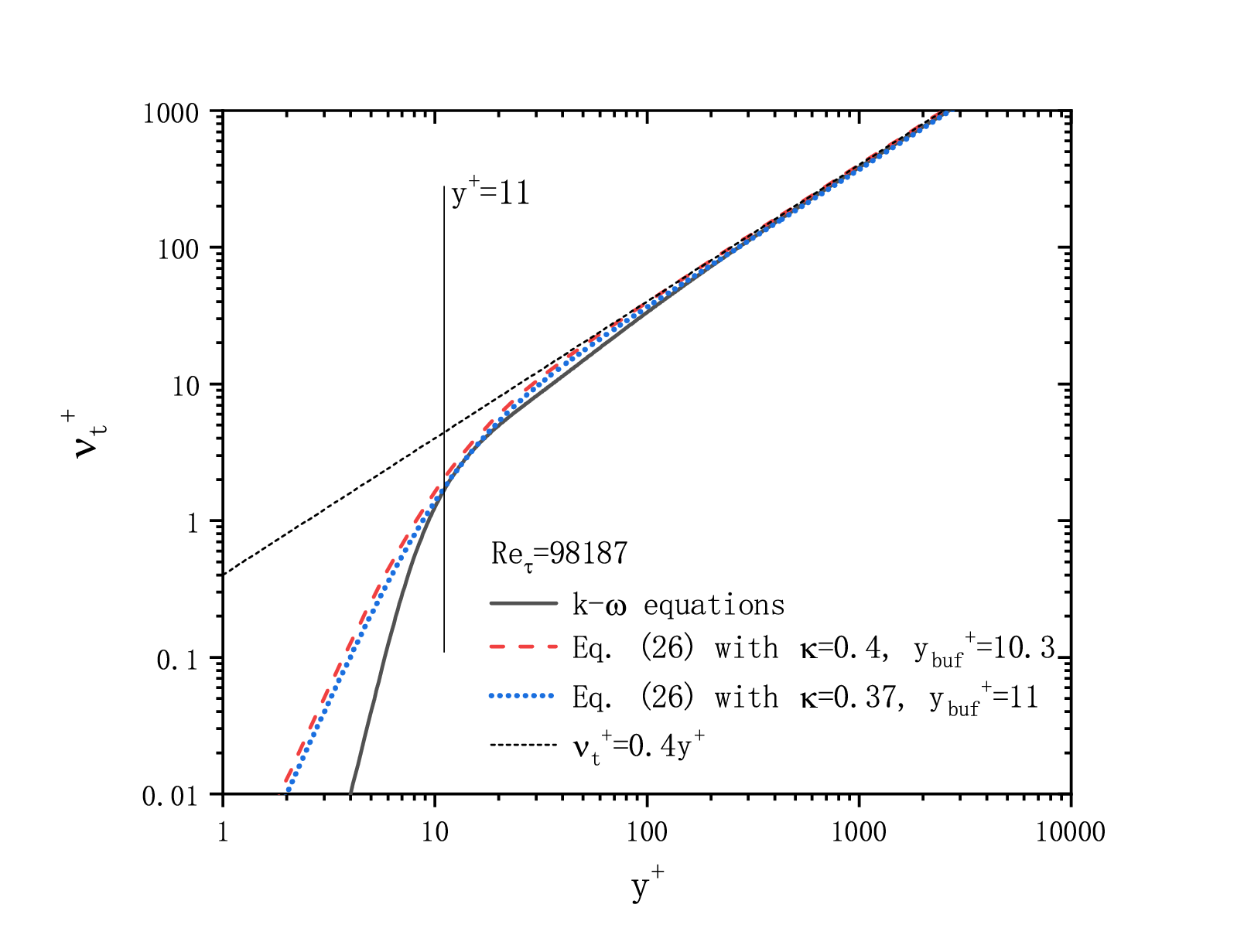}}}
\figcaption{(a) Validation of Eq. (\ref{eq:nut-ybuf-R}); (b) validation of Eq. (\ref{eq:nut-ybuf-ell}). The vertical line indicates the empirical location of $y_{buf}^+$.}\label{fig:nut-ybuf-valid}
\end{figure}

At first we propose an estimation for $y_{buf}^+$ by considering the leading-order deviation from the log-law when $y^+$ approaches $y_{buf}^+$ from the log-layer. In the log-layer, $R^+\approx\kappa y^+$ according to Eqs. (\ref{eq:k_log}) and (\ref{eq:omega_log}). Numerical simulations reveal that this approximation persists even at $y_{buf}^+$ -- the center of the buffer layer and log-layer. Thus,
\begin{align}
\label{eq:nut-ybuf-R} &\nu_t=\alpha^*R^+\approx \frac{\alpha_0^*R_k+\kappa y^+}{R_k+\kappa y^+}\kappa y^+.
\end{align}
On the other hand, $\tau^+=r\approx1$ and $\ell_{12}^+\gg1$ in the log-layer, such that $\nu_t^+\approx\ell_{12}^+$ according to (\ref{eq:Splus}) and (\ref{eq:nut+}). Close to $y_{buf}^+$, considering the approximated profile of $\ell_{12}^+$ with the leading-order deviation from the log-law, which can be derived from Eq. (\ref{eq:ell12formula}), one thus expects:
\begin{align}
\label{eq:nut-ybuf-ell} &\nu_t\approx\kappa y^+\left[1+\left(\frac{y_{buf}^+}{y^+}\right)^2\right]^{-5/4}.
\end{align}
Validation of (\ref{eq:nut-ybuf-R}) and (\ref{eq:nut-ybuf-ell}) is displayed in Fig. \ref{fig:nut-ybuf-valid} through comparing with the numerical simulation data. Indeed, the two approximated $\nu_t^+$ profiles are quite close to the numerical result at $y^+$ around and above $y_{buf}^+$. Then, at $y^+=y_{buf}^+$, we have
\begin{align}
\label{eq:ybuf-est} \frac{\alpha_0^*R_k+\kappa y_{buf}^+}{R_k+\kappa y_{buf}^+}\approx2^{-5/4},
\end{align}
which predicts $y_{buf}^+\approx10.3$, close to the empirical value of 11. Note that $y_{buf}^+$ and $\kappa$ can also be empirically estimated by using the numerical simulation data of $\ell_{12}^+$, which give $y_{buf}^+\approx11$ at all $Re_\tau$, but $\kappa\approx0.37$ at $Re_\tau=98187$, smaller than 0.4 set by the model, meaning that $\kappa$ is subject to a finite Reynolds number effect in the $k\mbox{-}\omega$ model. Eqs. (\ref{eq:nut-ybuf-R}) and (\ref{eq:nut-ybuf-ell}) with the empirical values of $\kappa$ and $y_{buf}^+$ are also plotted in Fig. \ref{fig:nut-ybuf-valid} for comparison. They are close to the theoretical profiles. Also plotted is the log-layer expression for $\nu_t^+$, which is $\nu_t^+=\kappa y^+$. The deviation of $\nu_t^+$ from the log-law is significant at $y_{buf}^+$, and has been fully described with the current leading-order approximations.

\noindent\textbf{B. Determining $y_{sub}^+$ from a viscous sublayer approximation}

Similarly we can determine $y_{sub}^+$ from a viscous sublayer approximation. In the viscous sublayer, $k$ and $\omega$ equations reduce to:
\begin{align}
\label{eq:k-sublayer} &\frac{100\beta_0^*\beta_0}{27}k^+\omega^+\approx\frac{d^2k^+}{dy^{+2}},\\[1mm]
\label{eq:omega-sublayer} &\beta \omega^{+2}\approx\frac{d^2\omega^+}{dy^{+2}},
\end{align}
whose solution, by applying the kinetic energy length, is
\begin{align}
\label{eq:sublayer-solu} &k^+=\frac{1}{2}\left(\frac{\kappa_ky_{ksub}^+}{y_{kbuf}^+}\right)^2y^{+2},\qquad \omega^+=\frac{6}{\beta y^{+2}}.
\end{align}
Thus,
\begin{align}
\label{eq:sublayer-R} &R^+=\frac{\beta}{12}\left(\frac{\kappa_ky_{ksub}^+}{y_{kbuf}^+}\right)^2y^{+4}.
\end{align}
Since $R^+\ll R_k$ in the viscous sublayer and $\alpha_0^* \ll 1$, from Eq. (\ref{eq:nut}) we have the approximation:
\begin{align}
\label{eq:sublayer-nut} &\nu_t^+\approx\alpha_0^*R^+\left(1+\frac{R^+}{\alpha_0^*R_k}\right).
\end{align}
On the other hand, because $S^+\approx1$ in the viscous sublayer, Eq. (\ref{eq:nut+}) reduces to:
\begin{align}
\label{eq:sublayer-nut-ell} &\nu_t^+\approx\ell_{12}^{+2},
\end{align}
where, according to (\ref{eq:ell12formula}),
\begin{align}
\label{eq:sublayer-ell} &\ell_{12}^+\approx\frac{\kappa y_{sub}^{+1.5}}{y_{buf}^{+2.5}}y^{+2}\left[1+\left(\frac{y^+}{y_{sub}^+}\right)^8\right]^{1.5/8}.
\end{align}
Therefore, from Eqs. (\ref{eq:sublayer-R})-(\ref{eq:sublayer-ell}), we obtain:
\begin{align}
\label{eq:sublayer-relation} &\frac{\beta\alpha_0^*}{12}\left(\frac{\kappa_ky_{ksub}^+}{y_{kbuf}^+}\right)^2
\left[1+\frac{\beta}{12\alpha_0^*R_k}\left(\frac{\kappa_ky_{ksub}^+}{y_{kbuf}^+}\right)^2y^{+4}\right]
\approx \frac{\kappa^2 y_{sub}^{+3}}{y_{buf}^{+5}}\left[1+\left(\frac{y^+}{y_{sub}^+}\right)^8\right]^{3/8}.
\end{align}
Letting $y^+$ approach zero in Eq. (\ref{eq:sublayer-relation}) leads to the wall constraint on the length functions:
\begin{align}
\label{eq:sublayer-constraint0} &\frac{\kappa^2 y_{sub}^{+3}y_{kbuf}^{+2}}{\kappa_k^2y_{ksub}^{+2}y_{buf}^{+5}}
=\frac{\beta\alpha_0^*}{12}.
\end{align}
Letting $y^+=y_{sub}^+$ in Eq. (\ref{eq:sublayer-relation}) leads to an estimation for $y_{sub}^+$:
\begin{align}
\label{eq:sublayer-constraint1} &\frac{\kappa^2 y_{sub}^{+7}}{\alpha_0^{*2}R_k y_{buf}^{+5}}=2^{3/8}-1.
\end{align}
In case of $\kappa=0.4$ and $y_{buf}^+=10.3$, (\ref{eq:sublayer-constraint1}) yields $y_{sub}^+\approx2.55$, which is quite close to the empirical value of 3. In case of using the empirical values of $\kappa$ and $y_{buf}^+$ (which are $\kappa=0.37$ and $y_{buf}^+=11$ at $Re_\tau=98187$), (\ref{eq:sublayer-constraint1}) yields $y_{sub}^+\approx2.74$, closer to the empirical value, indicating that taking into account a finite $Re$ effect would make the approximation better.

\noindent\textbf{C. Determining $y_{kbuf}^+$ from a log-layer approximation}

We proceed to estimate $y_{kbuf}^+$ by considering the leading-order deviation from the log-law when $y^+$ approaches $y_{kbuf}^+$ from the log-layer. In the log-layer, the production and dissipation terms balance in the $k$ equation, such that:
\begin{align}
\label{eq:balance-loglayer} &\frac{\beta^*k^+\omega^+}{S^+W^+}=\Theta,
\end{align}
where $\Theta\approx1$. When the buffer layer is approached, the turbulent convection plays a role, so $\Theta$ deviates from unity. In the SED theory $\Theta$ has been identified to possess similar multi-layer dilation symmetry as the length functions. Here, a two-layer formula is proposed for describing the variation of $\Theta$ from the log-layer to buffer layer, which reads:
\begin{align}
\label{eq:Theta} &\Theta=\left[1+\left(\frac{y_{kbuf}^+}{y^+}\right)^8\right]^{-1.2/8}.
\end{align}
Furthermore, since $\nu_t^+=\alpha^*R^+=W^+/S^+\approx(1-S^+)/S^+$, we have $S^+\approx(1+\alpha^*R^+)^{-1}$. Thus
\begin{align}
\label{eq:SW-approx} &S^+W^+\approx(1-S^+)S^+\approx \frac{\alpha^*R^+}{(1+\alpha^*R^+)^2}.
\end{align}
The dissipation term can be rewritten as
\begin{align}
\label{eq:diss-approx} &\beta^*k^+\omega^+=\beta^*k^{+2}/R^+.
\end{align}
Therefore,
\begin{align}
\label{eq:k-approx} &k^{+}\approx \sqrt{\Theta\frac{\alpha^*}{\beta^*}}\frac{R^+}{1+\alpha^*R^+}.
\end{align}
In Eqs. (\ref{eq:SW-approx})-(\ref{eq:k-approx}), and in $\alpha^*$ and $\beta^*$, $R^+$ is approximated with $R^+\approx \kappa y^+$.

On the other hand, according to Eq. (\ref{eq:kplus}), with $S^+\approx{\ell_{12}^+}^{-1}$ and $\ell_{12}^+$ and $\ell_k^+$ possessing the leading-order deviation from the log-law near $y_{kbuf}^+$, one has the approximation:
\begin{align}
\label{eq:k-ellk-loglayer} &k^+=\frac{1}{2}\ell_k^{+2}S^{+2}\approx\frac{1}{2}\frac{\ell_k^{+2}}{\ell_{12}^{+2}}
\approx\frac{1}{2}\frac{\kappa_k^2}{\kappa^2}\frac{\left[1+\left(y_{buf}^+/y^+\right)^2\right]^{5/2}}
{\left[1+\left(y_{kbuf}^+/y^+\right)^4\right]^{1/2}}.
\end{align}
In case of $y^+\gg y_{ybuf}^+$ in the log layer, (\ref{eq:k-approx}) and (\ref{eq:k-ellk-loglayer}) lead to:
\begin{align}
\label{eq:kappak} &\kappa_k=\sqrt{2}{\beta_0^*}^{-1/4}\kappa,
\end{align}
which yields $\kappa_k\approx1.03$. At $y^+= y_{ybuf}^+$, (\ref{eq:k-approx}) and (\ref{eq:k-ellk-loglayer}) lead to:
\begin{align}
\label{eq:kbuf-approx} &\left[1+\left(\frac{y_{buf}^+}{y_{kbuf}^+}\right)^2\right]^{5}
=2\Theta_{kbuf}\frac{\alpha_{kbuf}^*}{\beta_{kbuf}^*}\frac{\kappa^2y_{kbuf}^{+2}}{(1+\alpha_{kbuf}^*\kappa y_{kbuf}^+)^2},
\end{align}
where
\begin{align}
\label{eq:kbuf-approx-m} &\Theta_{kbuf}=2^{-1.2/8},\qquad \alpha_{kbuf}^*=\frac{\alpha_0^*+\kappa y_{kbuf}^+/R_k}{1+\kappa y_{kbuf}^+/R_k},\qquad
\beta_{kbuf}^*=\frac{100\beta_0/27+(\kappa y_{kbuf}^+/R_\beta)^4}{1+(\kappa y_{kbuf}^+/R_\beta)^4}.
\end{align}
In case of $\kappa=0.4$ and $y_{buf}^+=10.3$, Eq. (\ref{eq:kbuf-approx}) yields $y_{kbuf}^+=19.4$, which is only slightly larger than the empirical value of 17.5.

\begin{figure}[h]
\centering \mbox{ \subfigure[]{\includegraphics[width=75mm]{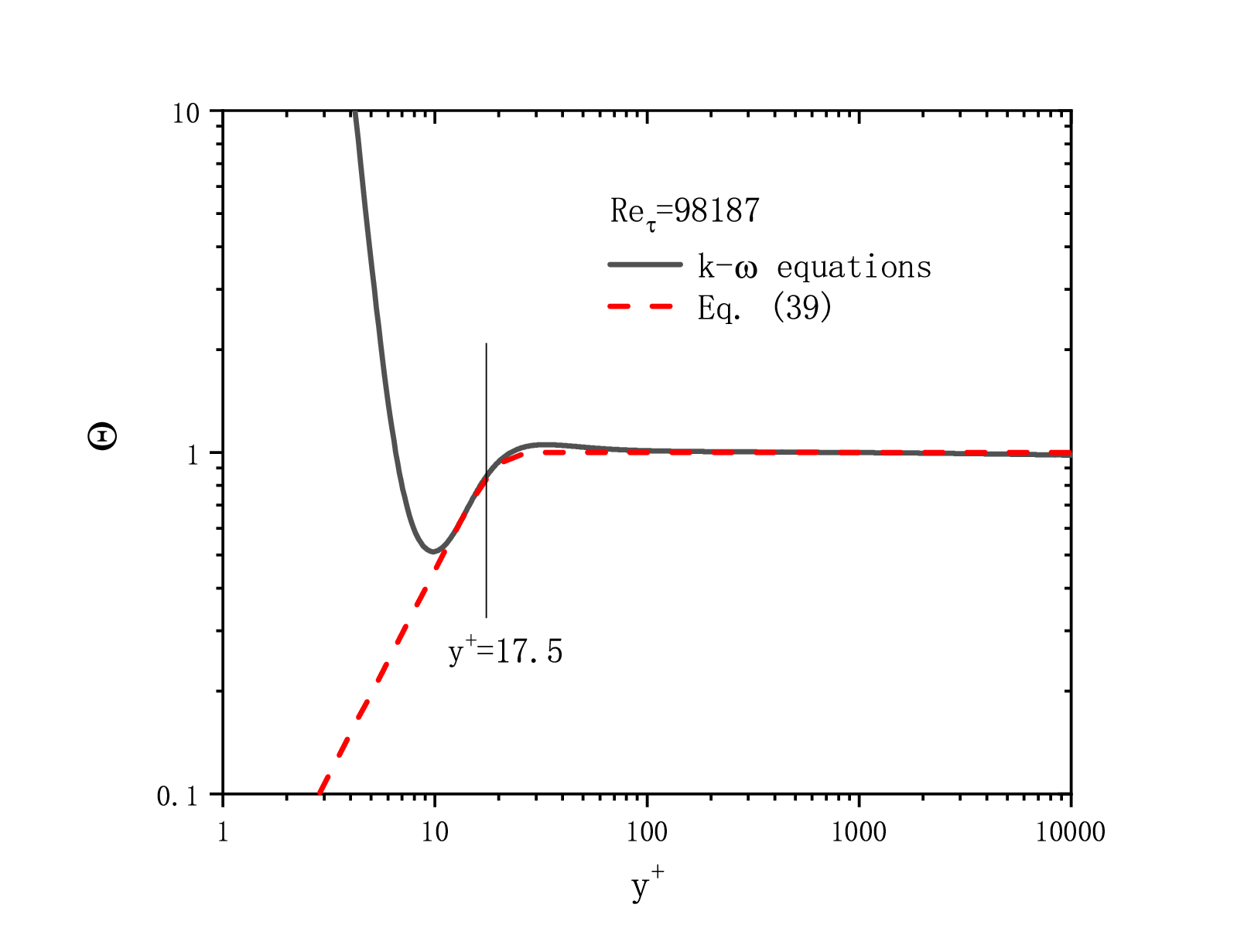}}\quad
\subfigure[]{\includegraphics[width=75mm]{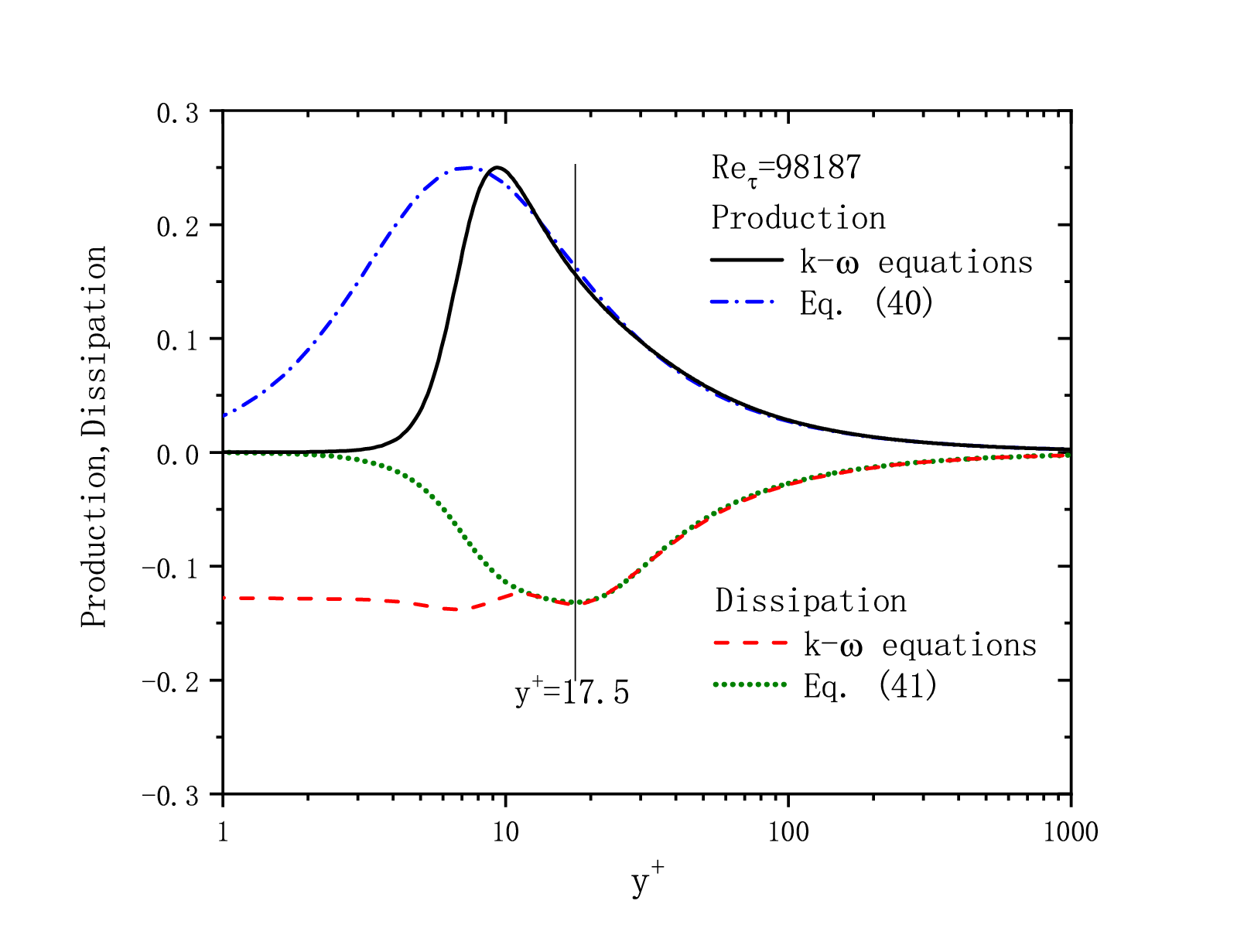}}}
\centering \mbox{ \subfigure[]{\includegraphics[width=75mm]{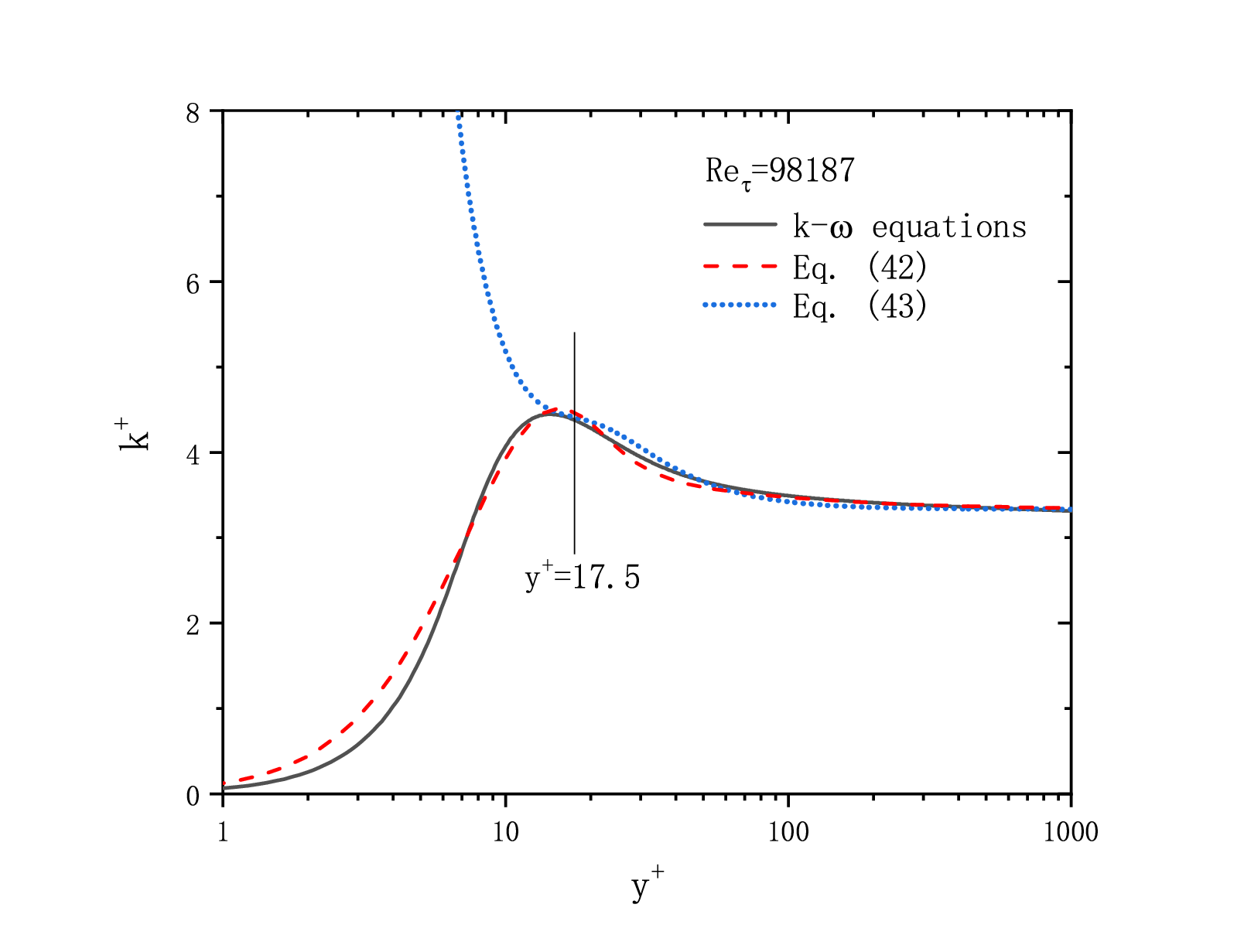}}}
\figcaption{Validation of (a) Eq. (\ref{eq:Theta}), (b) Eqs. (\ref{eq:SW-approx}) and (\ref{eq:diss-approx}); and (c) Eqs. (\ref{eq:k-approx}) and (\ref{eq:k-ellk-loglayer}). The vertical lines indicate the empirical location of $y_{kbuf}^+$.}\label{fig:prod-diss-valid}
\end{figure}

Validation of (\ref{eq:Theta})-(\ref{eq:k-ellk-loglayer}) is shown in Fig. \ref{fig:prod-diss-valid}. At $y^+$ around and above $y_{kbuf}^+$, (\ref{eq:Theta}) accurately characterizes the balance of production and dissipation in the $k$ equation, and (\ref{eq:SW-approx}) and (\ref{eq:diss-approx}) ideally describe the profiles of the production and dissipation terms in the $k$ equation. Fig. \ref{fig:prod-diss-valid}(c) validates Eqs. (\ref{eq:k-approx}) and (\ref{eq:k-ellk-loglayer}). The two approximations of $k$ are sufficiently accurate at around and above $y^+=y_{kbuf}^+$.

\noindent\textbf{D. Determining $y_{ksub}^+$ by using the wall constraint}

Now, $y_{ksub}^+$ can be calculated by using the wall constraint on the length functions, i.e. Eq. (\ref{eq:sublayer-constraint0}). By using the theoretical estimations, which are $\kappa=0.4$, $\kappa_k=1.03$, $y_{sub}^+=2.55$, $y_{buf}^+=10.3$, and $y_{kbuf}^+=19.4$, Eq. (\ref{eq:sublayer-constraint0}) predicts that $y_{ksub}^+=7.6$, which is comparable to the empirical estimation of $y_{ksub}^+=6$. Note that Eq. (\ref{eq:sublayer-constraint0}) is sensitive to errors owing to the the relatively large values of exponents in it, which come from the scaling exponents in the buffer layer, determined empirically in the current study. However, this sensitivity is not crucial to mean profiles since the viscous sublayer occupies a rather limited portion of boundary layer.

\noindent\textbf{E. Determining $r_c$ and $r_{kc}$ by local analysis around the pipe center}

In the pipe bulk the diffusion terms are neglected and the $k\mbox{-}\omega$ equations possess the following similar forms in the cylindrical coordinate\supercite{SED-k-omega}:
\begin{align}
\label{eq:k-bulk} &r^2-\beta_0^*k^{+2}+\nu_t^*\frac{1}{r}\frac{d}{dr}\left(r\sigma^*\nu_t^*\frac{dk^+}{dr}\right)=0,\\[1mm]
\label{eq:omega-bulk} &\alpha_\infty r^2-\beta k^{+2}+\nu_t^{*2}\frac{1}{r}\frac{d}{dr}\left(r\sigma\nu_t^*\frac{d\omega^*}{dr}\right)=0,
\end{align}
where $\nu_t^*=\nu_t^+/Re_\tau$ and $\omega^*=Re_\tau \omega^+$. Around the pipe center, dominant balance terms are the dissipation and turbulent convection, such that (\ref{eq:k-bulk}) and (\ref{eq:omega-bulk}), by using the leading order approximation, reduce to:
\begin{align}
\label{eq:k-bulk-approx}&\frac{1}{r}\frac{d}{dr}\left(r\nu_t^*\frac{dk^+}{dr}\right)\approx
\frac{\beta_0^*}{\sigma^*}\frac{k_0^{+2}}{\nu_{t0}^*},\\[1mm]
\label{eq:omega-bulk-approx} &\frac{1}{r}\frac{d}{dr}\left(r\nu_t^*\frac{d\omega^*}{dr}\right)\approx
\frac{\beta}{\sigma}\frac{k_0^{+2}}{\nu_{t0}^{*2}},
\end{align}
where $k_0^+$ and $\nu_{t0}^*$ are the corresponding values of $k^+$ and $\nu_t^*$ at the pipe center, respectively.

Now, we calculate the profiles of $k^+$ and $\omega^*$ through the length functions, which, in the pipe bulk, are written as:
\begin{align}
\label{eq:ell12-bulk} &\ell_{12}^+=\frac{\kappa Re_\tau}{mZ_c}(1-r^m)\left[1+\left(\frac{r_c}{r}\right)^2\right]^{1/4},\\[1mm]
\label{eq:ellk-bulk} &\ell_k^+=\frac{\kappa_k Re_\tau}{mZ_{kc}}
(1-r^m)\left[1+\left(\frac{r_{kc}}{r}\right)^2\right]^{1/2}.
\end{align}
Around the pipe center, $S^+=\sqrt{W^+}/\ell_{12}^+\approx \sqrt{r}/\ell_{12}^+$. Consequently,
\begin{align}
\label{eq:nu-pipecenter} &\nu_t^*=\frac{1}{Re_\tau}\ell_{12}^{+2}S^+\approx \nu_{t0}^*(1-r^m)\left[1+\left(\frac{r}{r_c}\right)^2\right]^{1/4},\\[1mm]
\label{eq:k-pipecenter} &k^+=\frac{1}{2}\ell_k^{+2}S^{+2}\approx
k_0^+\left[1+\left(\frac{r}{r_{kc}}\right)^2\right]\left[1+\left(\frac{r}{r_{c}}\right)^2\right]^{-1/2}\\
\label{eq:omega-center} &\omega^*=Re_\tau \alpha^*\frac{k^+}{\nu_t^+}\approx \frac{k_0^+}{\nu_{t0}^*}(1-r^m)^{-1}
\left[1+\left(\frac{r}{r_{kc}}\right)^2\right]\left[1+\left(\frac{r}{r_{c}}\right)^2\right]^{-3/4},
\end{align}
where $\nu_{t0}^*=\frac{\kappa \sqrt{r_c}}{mZ_c}$, $k_0^+=\frac{1}{2}\left(\frac{\kappa_k Z_c r_{kc}}{\kappa Z_{kc} \sqrt{r_{c}}}\right)^2$. Using the above approximations to substitute for the corresponding terms in Eqs. (\ref{eq:k-bulk-approx}) and (\ref{eq:omega-bulk-approx}) and assuming $m=2$ (determined below), we derive:
\begin{align}
\label{eq:constraint-center1} &\frac{k_0^+r_c^2r_{kc}^2}{2\nu_{t0}^{*2}(2r_c^2-r_{kc}^2)}=\frac{\sigma^*}{\beta_0^*}\approx 6.67,\\
\label{eq:constraint-center2} &\frac{k_0^+r_c^2r_{kc}^2}{\nu_{t0}^{*2}(4r_c^2r_{kc}^2+4r_c^2-3r_{kc}^2)}=\frac{\sigma}{\beta}\approx 7.06.
\end{align}
Solving (\ref{eq:constraint-center1}) and (\ref{eq:constraint-center2}) with $\kappa=0.4$ and $\kappa_k=1.03$ gives $r_c=0.45$ and $r_{kc}=0.43$, which are somehow larger than the empirical values of 0.4 and 0.35. In case of assuming $r_c=r_{kc}$ and solving $r_c$ with Eq. (\ref{eq:constraint-center1}) solely (since $k$-equation is more physical), an explicit expression can be derived for $r_c$:
\begin{align}
\label{eq:constraint-center3} &r_c=r_{kc}=\sqrt{\frac{\sigma^*}{2\sqrt{\beta_0^*}}}\kappa=0.4,
\end{align}
closer to the empirical value, where Eq. (\ref{eq:kappak}) is employed.

Here we end the theoretical analysis for determining the multi-layer structure parameters in Eqs. (\ref{eq:ell12formula}) and (\ref{eq:ellkformula}). In Table \ref{tab1:parameters}, the theoretical parameter values are concluded and compared with the corresponding empirical values measured through directly fitting the computed profiles of $\ell_{12}^+$ and $\ell_k^+$.

\subsection{Predicting the critical locations of the transition functions via the analytic solution}\label{subsec:transition prediction}

There are three transition functions (Eqs. (\ref{eq:Rbeta})-(\ref{eq:Romega})) in the $k\mbox{-}\omega$ model equations, aiming to control the transition from the near-wall flow to the outer-layer flow. Constructing analytic relationships between this setting and the predicted multi-layer structure of TBL is crucial for understanding and improving the $k\mbox{-}\omega$ model. Since the transitions occur in the buffer layer regime where all terms in the balance equations play roles, conventional methods like the perturbation analysis are restricted. Here we derive more relations between the model parameters and the multi-layer structure parameters by studying the critical locations of the transition functions with the analytic solution.

The first is with respect to $\beta^*$, in which a critical $R^+$ ($=R_\beta$) is set by Wilcox for adjusting the magnitude of kinetic energy dissipation from viscous sublayer to outer layer. Let $y_\beta^+$ denotes the location where $R^+=R_\beta$. Then, at $y_\beta^+$, we have:
\begin{align}
\label{eq:nut-Rbeta-approx} &\nu_t^+=\alpha_\beta^*R_\beta\approx\frac{1-S(y_\beta^+)}{S(y_\beta^+)},
\end{align}
where $\alpha_\beta^*=\frac{\alpha_0^*+R_\beta/R_k}{1+R_\beta/R_k}$. In other words,
\begin{align}
\label{eq:S-Rbeta-approx} &S(y_\beta^+)=\frac{1}{1+\alpha_\beta^*R_\beta}.
\end{align}
Since $\ell_{12}^+\approx \sqrt{1-S^+}/S^+$ in the buffer layer and above,
\begin{align}
\label{eq:ell12-Rbeta-approx} &\ell_{12}^+(y_\beta^+)=\sqrt{(1+\alpha_\beta^*R_\beta)\alpha_\beta^*R_\beta}.
\end{align}
On the other hand, according to Eq. (\ref{eq:ell12formula}), $\ell_{12}^+$ at $y_\beta^+$ can be approximated with
\begin{align}
\label{eq:ell12-Rbeta-approx1} &\ell_{12}^+(y_\beta^+)\approx\kappa y_\beta^+\left(\frac{y_\beta^+}{y_{buf}^+}\right)^{2.5}\left[1+\left(\frac{y_\beta^+}{y_{buf}^+}\right)^2\right]^{-2.5/2}.
\end{align}
(\ref{eq:ell12-Rbeta-approx}) and (\ref{eq:ell12-Rbeta-approx1}) establish a relationship for calculating $y_\beta^+$ for a given $R_\beta$, in which $y_{buf}^+$ can be estimated by using Eq. (\ref{eq:ybuf-est}). For example, taking the theoretical values of $\kappa$ and $y_{buf}^+$, in case of $R_\beta=8$ as the model setting, we derive $y_\beta^+\approx 18.2$, which is quite close to the measured value of 19 at $Re_\tau=98198$.

The second is with respect to $\alpha^*$, which is the coefficient in the constitutive relation for adjusting the wall-normal variation of eddy viscosity. The analysis is similar. A critical $R^+$ ($=R_k$) is set for $\alpha^*$, at which the location is denoted with $y_k^+$. Then, at $y_k^+$, we have:
\begin{align}
\label{eq:nut-Rk-approx} &\nu_t^+=\alpha_k^*R_k\approx\frac{1-S(y_k^+)}{S(y_k^+)},
\end{align}
where $\alpha_k^*=(1+\alpha_0^*)/2$. Consequently,
\begin{align}
\label{eq:S-Rk-approx} &S(y_k^+)=\frac{1}{1+\alpha_k^*R_k}=\frac{2}{2+R_k(1+\alpha_0^*)}.
\end{align}
Following a similar procedure as above, we can obtain the following relation:
\begin{align}
\label{eq:ell12-Rk-approx} &\ell_{12}^+(y_k^+)=\frac{1}{2}\sqrt{R_k(1+\alpha_0^*)[2+R_k(1+\alpha_0^*)]}
\approx\kappa y_k^+\left(\frac{y_k^+}{y_{buf}^+}\right)^{2.5}\left[1+\left(\frac{y_k^+}{y_{buf}^+}\right)^2\right]^{-2.5/2}.
\end{align}
(\ref{eq:ell12-Rk-approx}) is a relation for calculating $y_k^+$ for a given $R_k$. For example, taking the theoretical values of $\kappa$ and $y_{buf}^+$, in case of $R_k=6$ as the model setting, we derive $y_k^+\approx 14.6$, which is quite close to the measured value of 15.2.

The third transition function is with respect to $\alpha$, which, with $\alpha^*$ together, adjusts the variation of the production and thus the balance of $\omega$. A critical $R^+$ ($=R_\omega$) is set for the transition function $\alpha\alpha^*$, at which we assume $y^+=y_\omega^+$. Then, following a similar procedure as in the derivation of $y_\beta^+$, we have,
\begin{align}
\label{eq:ell12-Romega-approx} &\ell_{12}^+(y_\omega^+)=\sqrt{(1+\alpha_\omega^*R_\omega)\alpha_\omega^*R_\omega}\approx
\kappa y_\omega^+\left(\frac{y_\omega^+}{y_{buf}^+}\right)^{2.5}\left[1+\left(\frac{y_\omega^+}{y_{buf}^+}\right)^2\right]^{-2.5/2},
\end{align}
where $\alpha_\omega^*=\frac{\alpha_0^*+R_\omega/R_k}{1+R_\omega/R_k}$. By using the theoretical values of $\kappa$, $y_{sub}^+$ and $y_{buf}^+$, Eq. (\ref{eq:ell12-Romega-approx}) predicts that $y_\omega^+=8.9$, which is exactly the same as the measured value.

Note that the above predictions to the critical transition locations as well as the multi-layer structure parameters not only validate the analytic solutions of the length functions, but also construct relationships between the model parameters in the $k\mbox{-}\omega$ equations and the multi-layer structure parameters of predicted flows. The latter is crucial in that, to gain better predictions, engineers may want to tune the model parameters, which has remained implicit for decades, but now made explicit.

\subsection{Validating the multi-layer analytic solution}\label{subsec:validation}

Herein we validate the analytic solutions with the numerical simulation results of the $k\mbox{-}\omega$ model equations computed by using the code presented in Wilcox's book. We take three $Re_\tau$ for comparisons, which possess direct numerical simulation\supercite{WuXH-DNS} (at $Re_\tau=1142$) and experimental data\supercite{SuperPipe1998,SuperPipe2012} (at $Re_\tau=98187$ and $528550$), and cover a considerably wide range of $Re_\tau$. The parameters used for calculating the length functions (i.e. Eqs. (\ref{eq:ell12formula}) and (\ref{eq:ellkformula})) are listed in Table \ref{tab1:parameters} for each $Re_\tau$. The analytic solutions of the $k\mbox{-}\omega$ model equations are calculated accordingly via Eqs. (\ref{eq:Splus})-(\ref{eq:omega+}).

\begin{figure}[h]
\centering \mbox{ \subfigure[]{\includegraphics[width=75mm]{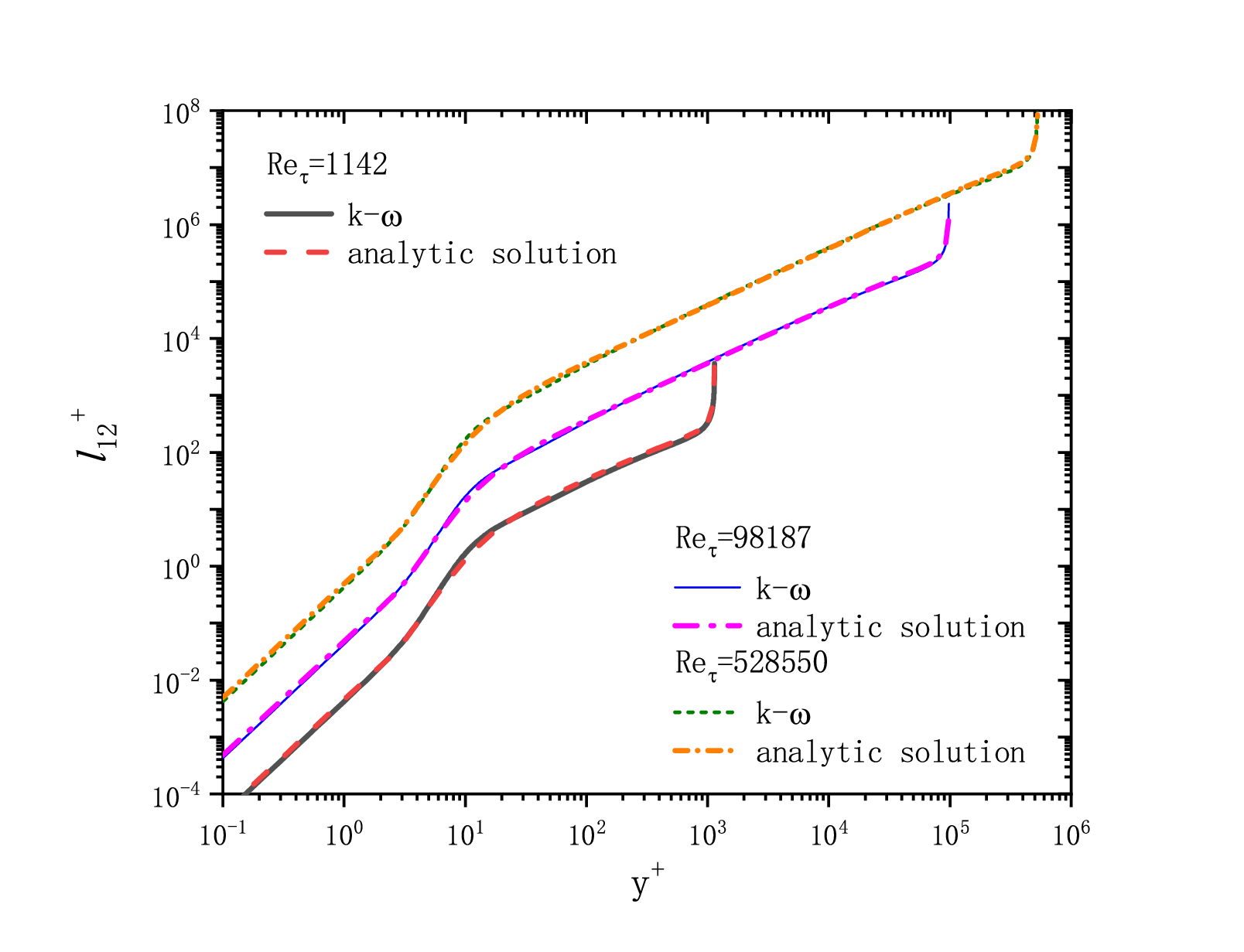}}\quad
\subfigure[]{\includegraphics[width=75mm]{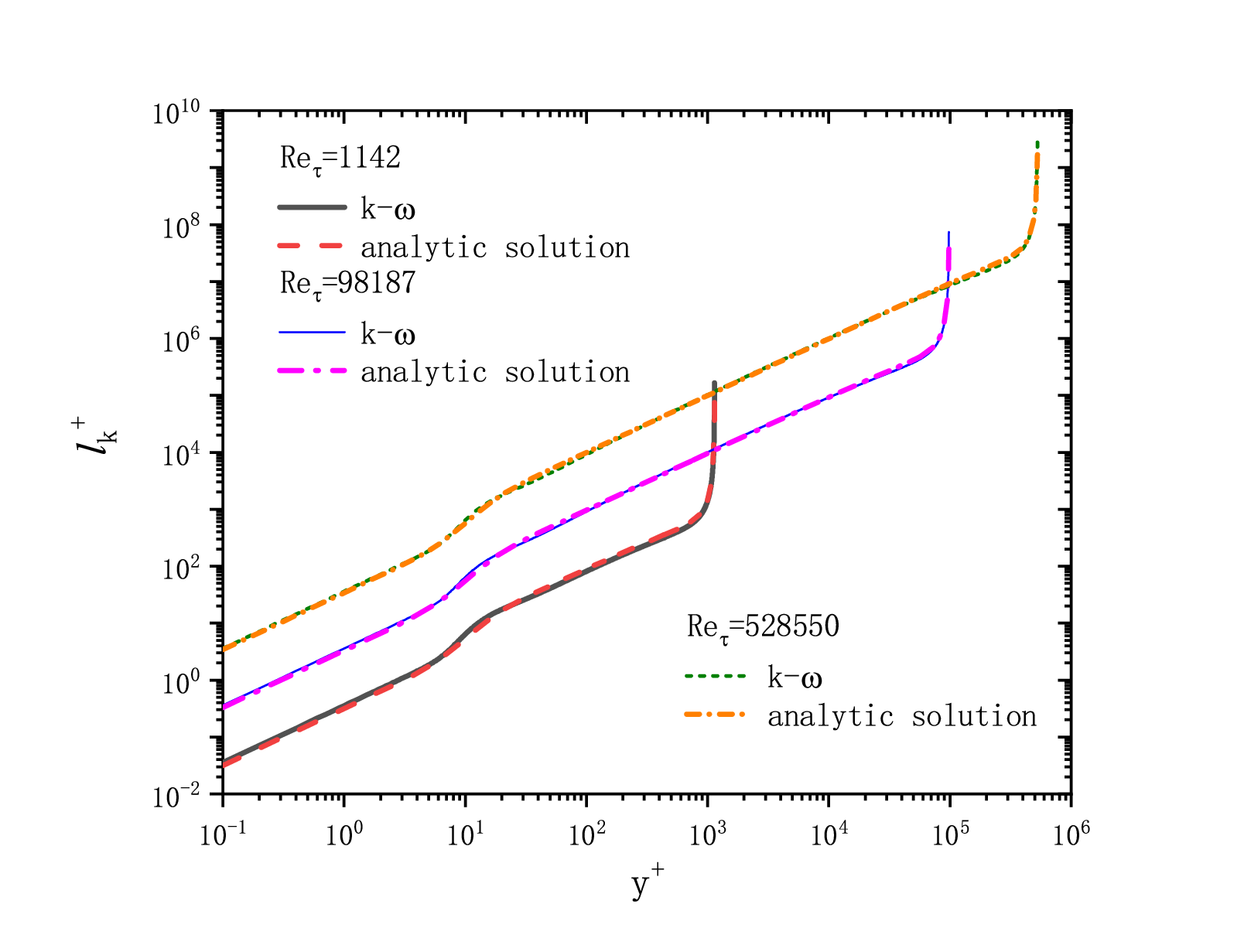}}}
\figcaption{Comparisons between the numerical simulation results of the $k\mbox{-}\omega$ model equations and the predictions of the analytic solutions for (a) $\ell_{12}^+$, and (b) $\ell_{k}^+$. To display clearly the profiles are vertically shifted for $Re_\tau=98187$ and $528550$.}\label{fig:ell12-ellk-valid}
\end{figure}

Figure \ref{fig:ell12-ellk-valid} compares the analytic solutions of $\ell_{12}^+$ and $\ell_{k}^+$ with the numerical simulation data. The theoretical formulas excellently describe the numerically-measured profiles for both $\ell_{12}^+$ and $\ell_{k}^+$ over the whole pipe radius. Note that the empirical parameters in Table \ref{tab1:parameters} are used to for calculating $\ell_{12}^+$ and $\ell_{k}^+$. This is not a defect to the current theoretical analysis because only $\kappa$ and $\kappa_k$ are varied, which is necessary owing to the finite Reynolds number effect in the $k\mbox{-}\omega$ model.

\begin{figure}[h]
\centering \includegraphics[width=100mm]{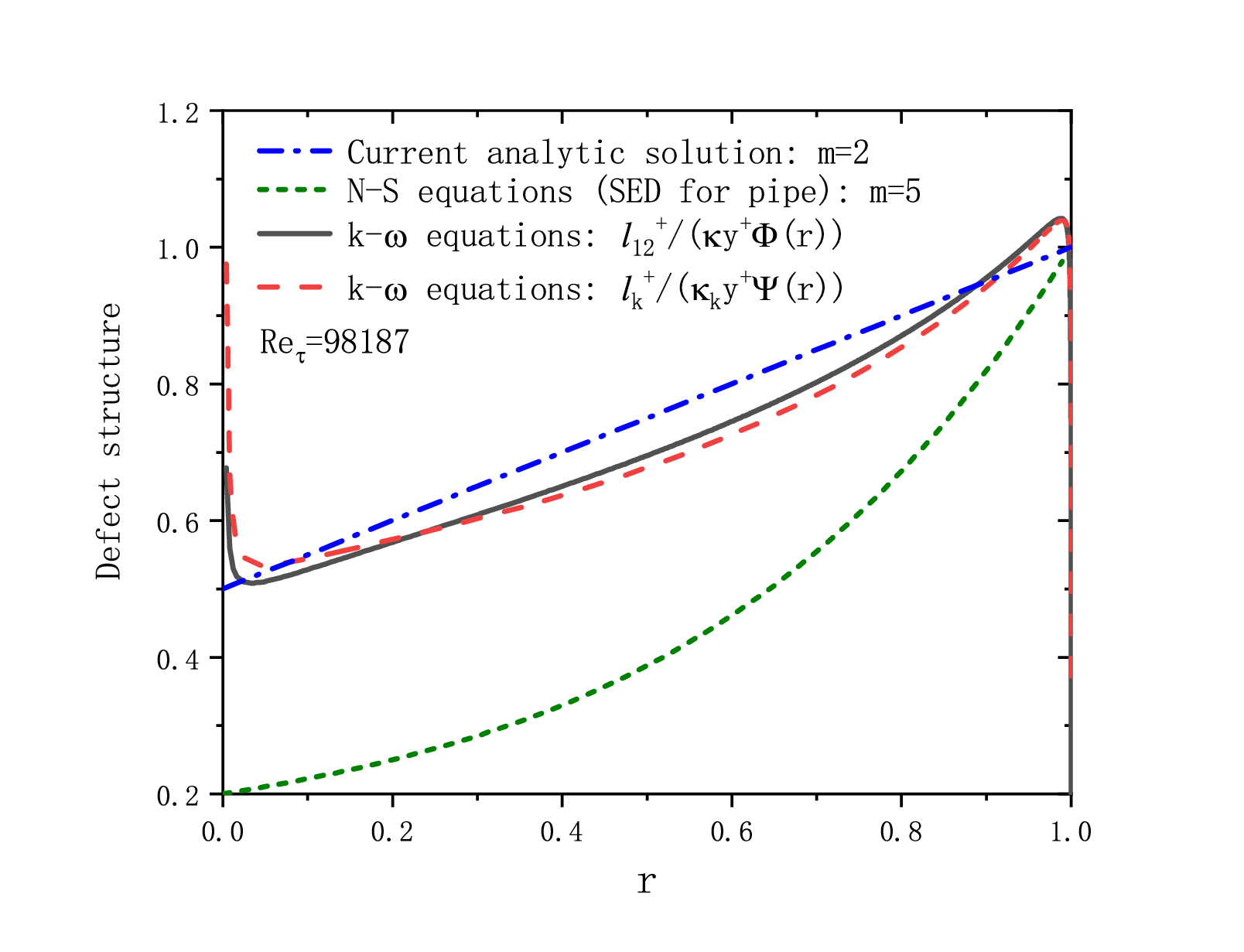}
\figcaption{Validation of the defect structure: $\frac{1-r^m}{m(1-r)}$. $m=2$ in the current analytic solution and $m=5$ in the SED theory for pipe flow. The data of the $k\mbox{-}\omega$ equations are calculated by the measured $\ell_{12}^+$ and $\ell_k^+$ compensated with the log-law and the center-core structure in the current analytic solution.}\label{fig:defect-valid}
\end{figure}

Figure \ref{fig:ell12-ellk-valid} reveals a difference between the true pipe flow (predicted with the N-S equations and described by the SED theory) and the $k\mbox{-}\omega$-predicted one. In the $k\mbox{-}\omega$-predicted flow, the log-layer that is supposed to be the overlap region between the inner flow and the bulk, extends far towards the pipe center. In other words, there lacks a sufficient plateau for $\ell_{12}^+$ and $\ell_{k}^+$ (or pit for the corresponding diagnostic functions) in the bulk of the $k\mbox{-}\omega$-predicted flow. This plateau is described with the defect structure (i.e. $1-r^m$) in the multi-layer structure of length functions. Here we present a validation of the defect structure by plotting:
\begin{align}
\label{eq:defect-law} &\frac{1-r^m}{m(1-r)}=\frac{\ell_{12}^+}{\kappa y^+ \Phi (r)}=\frac{\ell_k^+}{\kappa_k y^+ \Psi (r)},
\end{align}
where $\Phi (r)$ and $\Psi (r)$ denote the center-core structures, e.g. $\Phi (r)=\frac{1}{Z_c}\left[1+\left(\frac{r_c}{r}\right)^2\right]^{1/4}$, and $\Psi (r)=\frac{1}{Z_{kc}}\left[1+\left(\frac{r_{kc}}{r}\right)^2\right]^{1/4}$, respectively. Note that the left hand side of (\ref{eq:defect-law}) equals $1/m$ when $r$ approaches zero and equals 1 when $r$ approaches unity. As indicated in Fig. \ref{fig:defect-valid}, the $k\mbox{-}\omega$ model equations possess a mild defect structure with $m\approx2$ (note $m>1$, in order for a defect structure to occur), which is smaller than $m=5$ for the SED-determined scaling for the N-S turbulent pipe flows. Note that this difference in $m$ is consistent with the observation of  different pits in the diagnostic functions of $d\log\ell_{12}^+/d\log y^+$, and $d\log\ell_{k}^+/d\log y^+$, as shown in Fig. \ref{fig:dlogdlog}. This difference has been a target for an improvement by the SED $k\mbox{-}\omega$ model\supercite{SED-k-omega,ChenX-WeiBB} explained below.

\begin{center}
\abovecaptionskip 0pt \belowcaptionskip 1pt
\renewcommand{\arraystretch}{1.2}
{\scriptsize \tabcaption{Empirical and theoretical multi-layer structure parameters in the $k\mbox{-}\omega$ model-predicted pipe flows described with Eqs. (\ref{eq:ell12formula}) and (\ref{eq:ellkformula}) at different $Re_\tau$. In (\ref{eq:ell12formula}) and (\ref{eq:ellkformula}), $\gamma_w=1.5$ and $\gamma_k=1$. Listed also are the multi-layer structure parameters in the pipe flow at $Re_\tau=528550$ predicted by the SED $k\mbox{-}\omega$ model\supercite{SED-k-omega}, in which $\gamma_w=1.5$ and $\gamma_k=1$., and the multi-layer structure parameters in the SED theory for pipe flow\supercite{SEDPart2}, in which $\gamma_w=0.5$ and $\gamma_k=0.5$.} \label{tab1:parameters}
\noindent\begin{tabular*}{\textwidth}{@{\extracolsep{\fill}}@{~~}cccccccccc}
\toprule%
$Re_\tau$ & $\kappa$ & $\kappa_k$ & $y_{sub}^+$ & $y_{buf}^+$ & $y_{ksub}^+$ & $y_{kbuf}^+$ & $r_c$ & $r_{kc}$ & $m$\\
\midrule%
1142 & 0.35 & 0.92 & 3 & 11  & 6 & 17.5 & 0.4 & 0.35 & 2\\
98187 & 0.37 & 0.96 & 3 & 11  & 6 & 17.5 & 0.4 & 0.35 & 2 \\
528550 & 0.38 & 0.99 & 3 & 11  & 6 & 17.5 & 0.4 & 0.35 & 2 \\
Current theory for $k\mbox{-}\omega$ & 0.4 & 1.03 & 2.55 & 10.3  & 7.6 & 19.4 & 0.4(0.45) & 0.4(0.43) & 2 \\
SED $k\mbox{-}\omega$ at $Re_\tau=528550$ & 0.42 & 1.09 & 3 & 13 & 6 & 21.5 & 0.3 & 0.3 & 5 \\
SED theory for pipe & 0.45 & 1.02 & 9.7 & 41 & 9.7 & 41 & 0.27 & 0.27 & 5 \\

\bottomrule
\end{tabular*}
\small }\\[4mm]
\end{center}

\begin{figure}[h]
\centering \mbox{ \subfigure[]{\includegraphics[width=75mm]{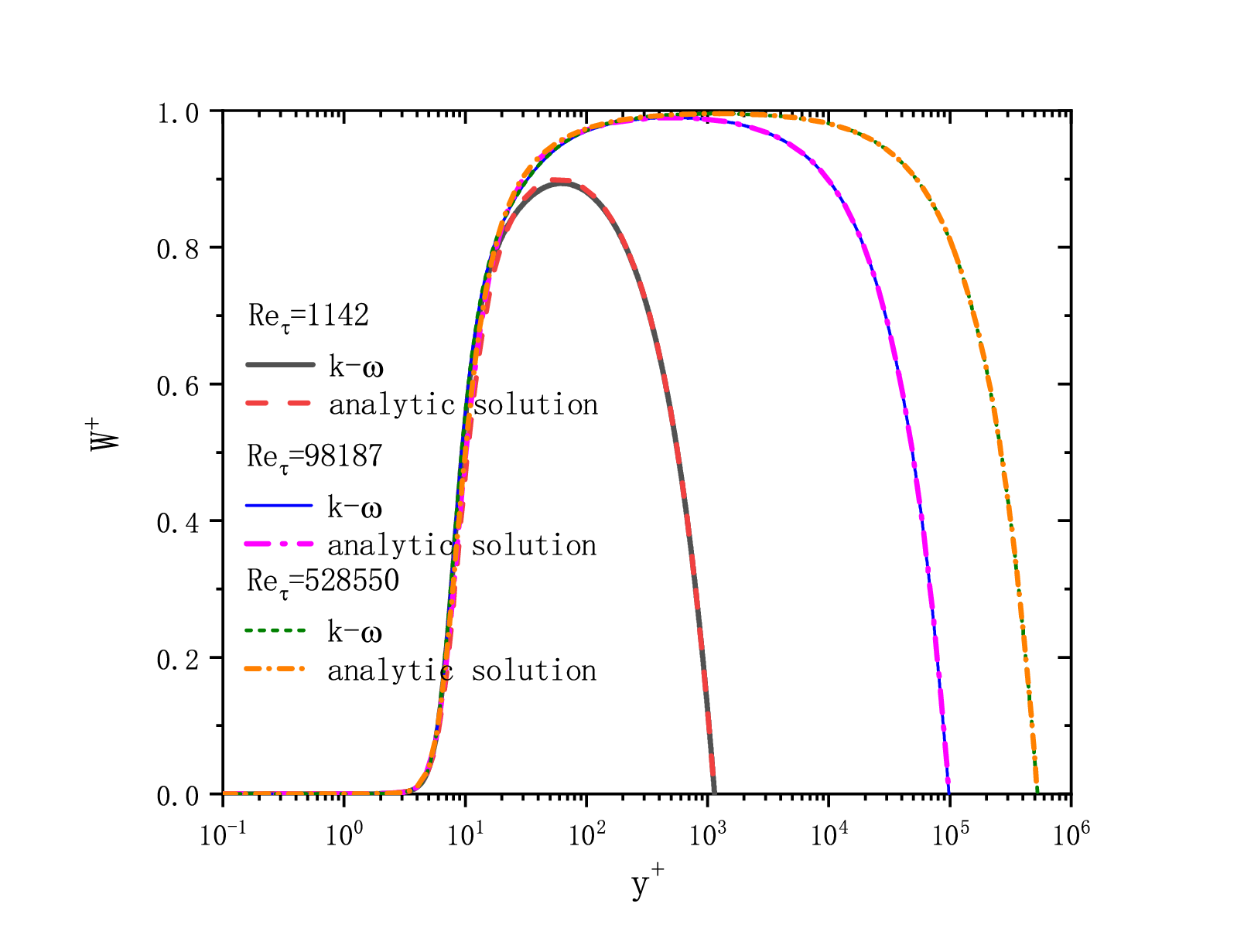}}\quad
\subfigure[]{\includegraphics[width=75mm]{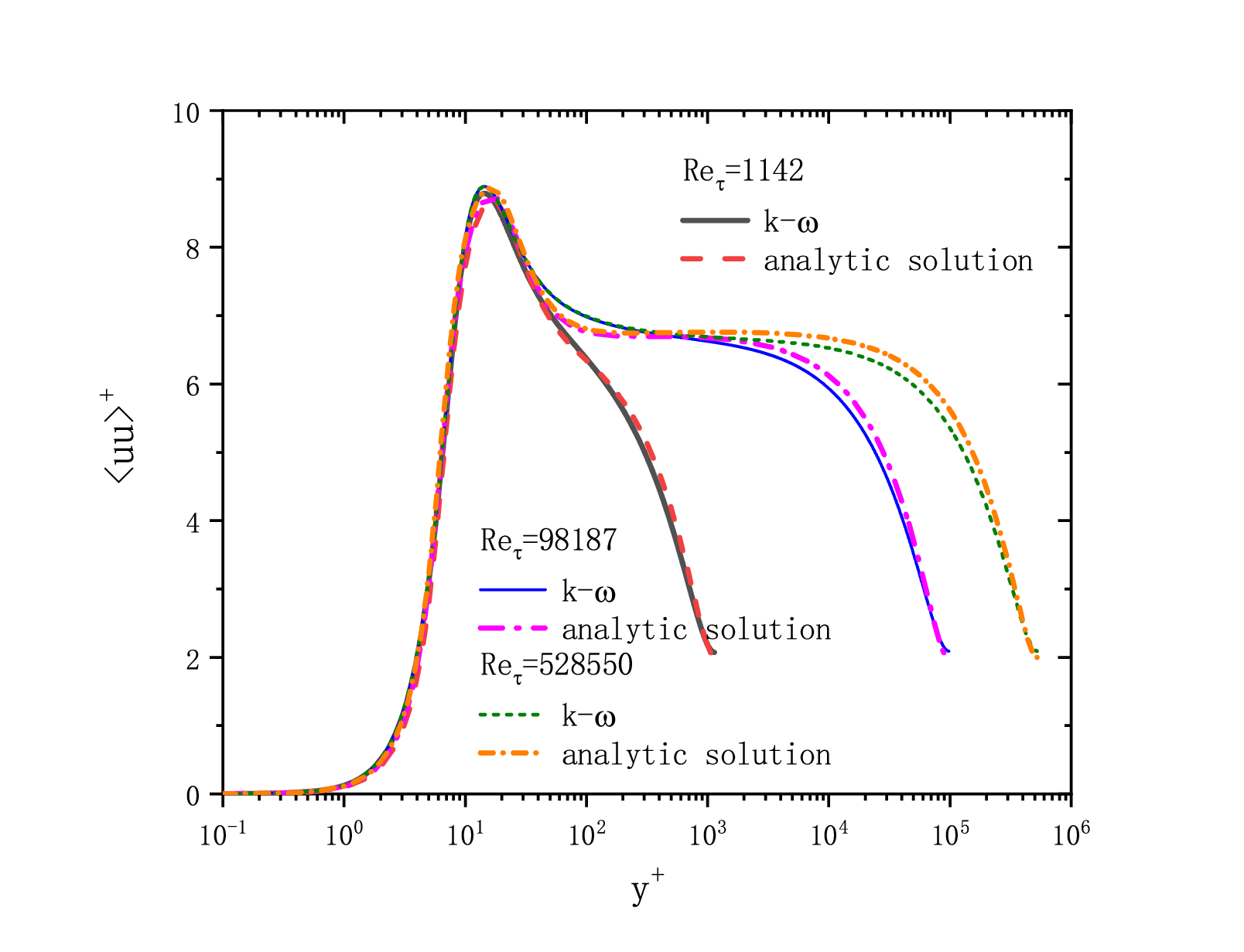}}}
\figcaption{Comparisons between the numerical simulation results of the $k\mbox{-}\omega$ model equations and the predictions of the analytic solutions for (a) $W^+$, and (b) ${\left \langle uu \right \rangle}^+$ ($=2k^+$).}\label{fig:W-uu-valid}
\end{figure}

$W^+$ and ${\left \langle uu \right \rangle}^+$ ($=2k^+$) predicted by the analytic solution are compared with the numerical results in Fig. \ref{fig:W-uu-valid}. Again, the theoretical predictions agree with the numerical simulations quite well. \textcolor{black}{Minor differences can be found for the ${\left \langle uu \right \rangle}^+$ profiles above the buffer layer and beyond the log-layer. As is also shown in Fig. \ref{fig:dlogdlog}(b), an additional undulation exists in this regime for $d\log\ell_{k}^+/d\log y^+$ of the $k\mbox{-}\omega$ model (owing to the unique characteristic of the model), in contrast to that of the SED profile, which reveals a special complexity in the multi-layer structure of $k$ for the $k\mbox{-}\omega$ model. A remedy of this difference is easy. One only needs to introduce another layer to describe this abnormal scaling law in the regime, by using the same universal dilation ansatz of the SED. However, it is only a second-order effect that can be neglected (if one compares this difference with the significant deviation of the $k\mbox{-}\omega$-predicted $k$ profile from the experimental data, as shown in Fig. \ref{fig:SEDkw-U-k-valid}(b)). For simplicity, it is not considered in the current study.}

\begin{figure}[h]
\centering \mbox{ \subfigure[]{\includegraphics[width=75mm]{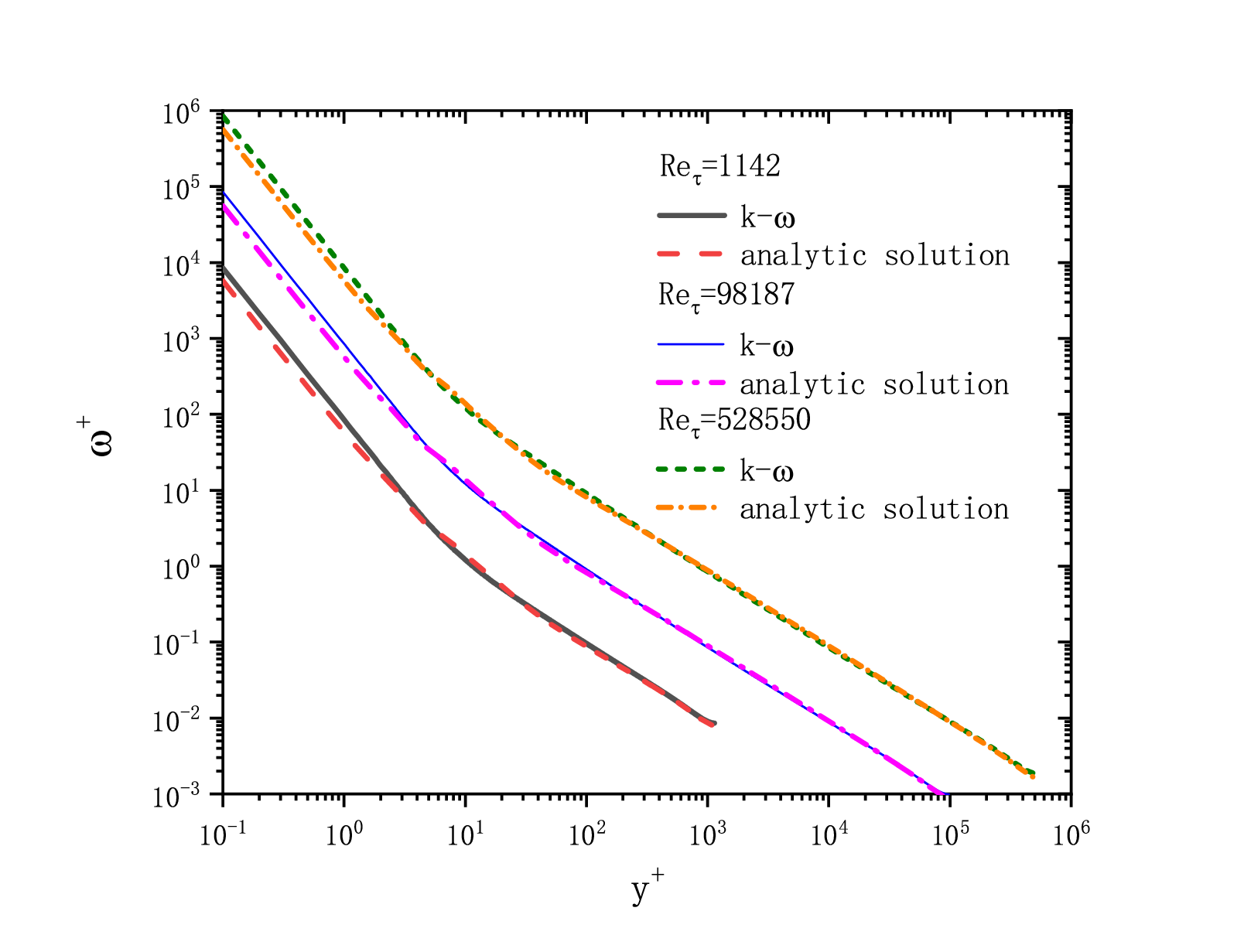}}\quad
\subfigure[]{\includegraphics[width=75mm]{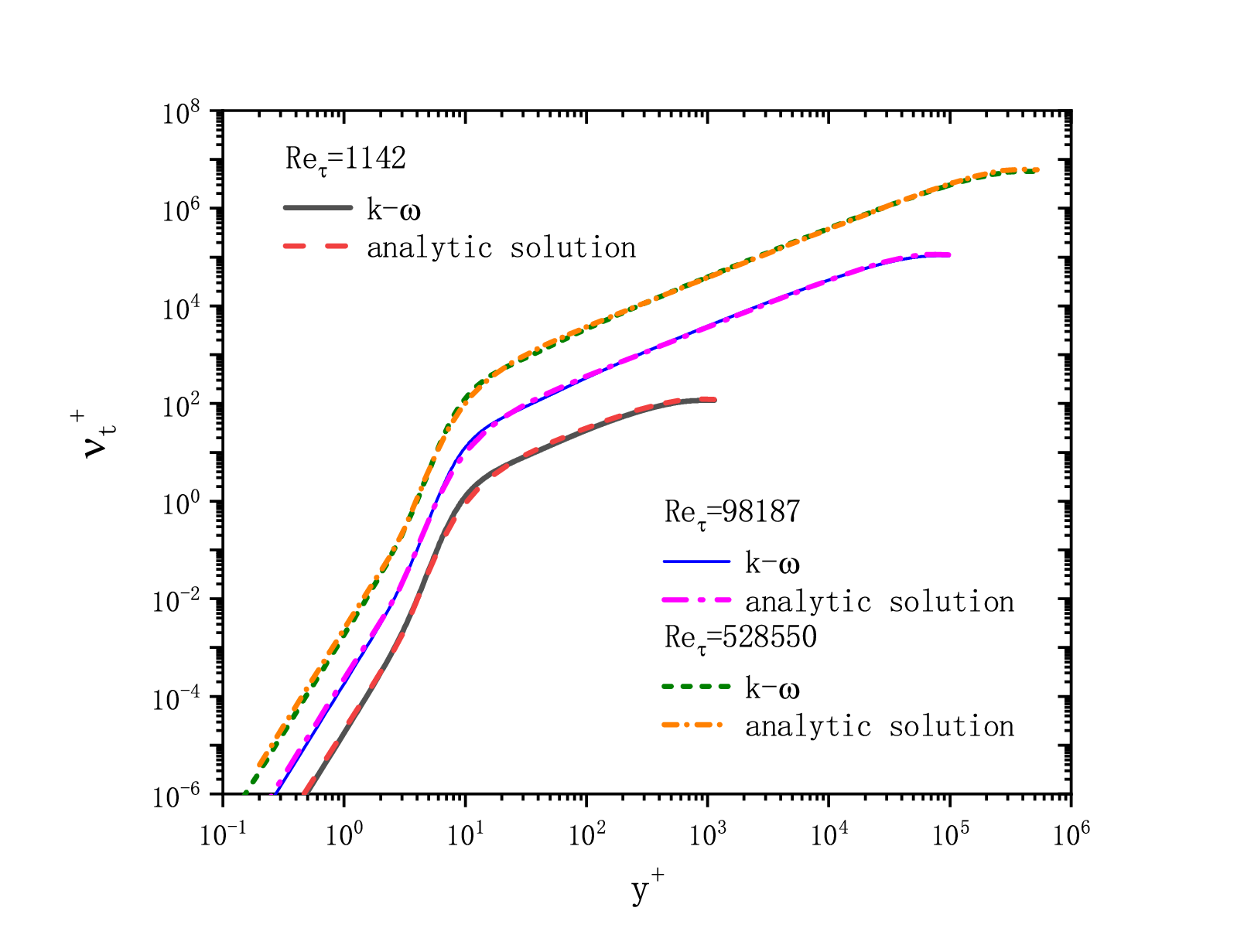}}}
\figcaption{Comparisons between the numerical simulation results of the $k\mbox{-}\omega$ model equations and the predictions of the analytic solutions for (a) $\omega^+$, and (b) $\nu_t^+$. To display clearly the profiles are vertically shifted for $Re_\tau=98187$ and $528550$.}\label{fig:omega-nut-valid}
\end{figure}

\begin{figure}[h]
\centering \mbox{ \subfigure[]{\includegraphics[width=75mm]{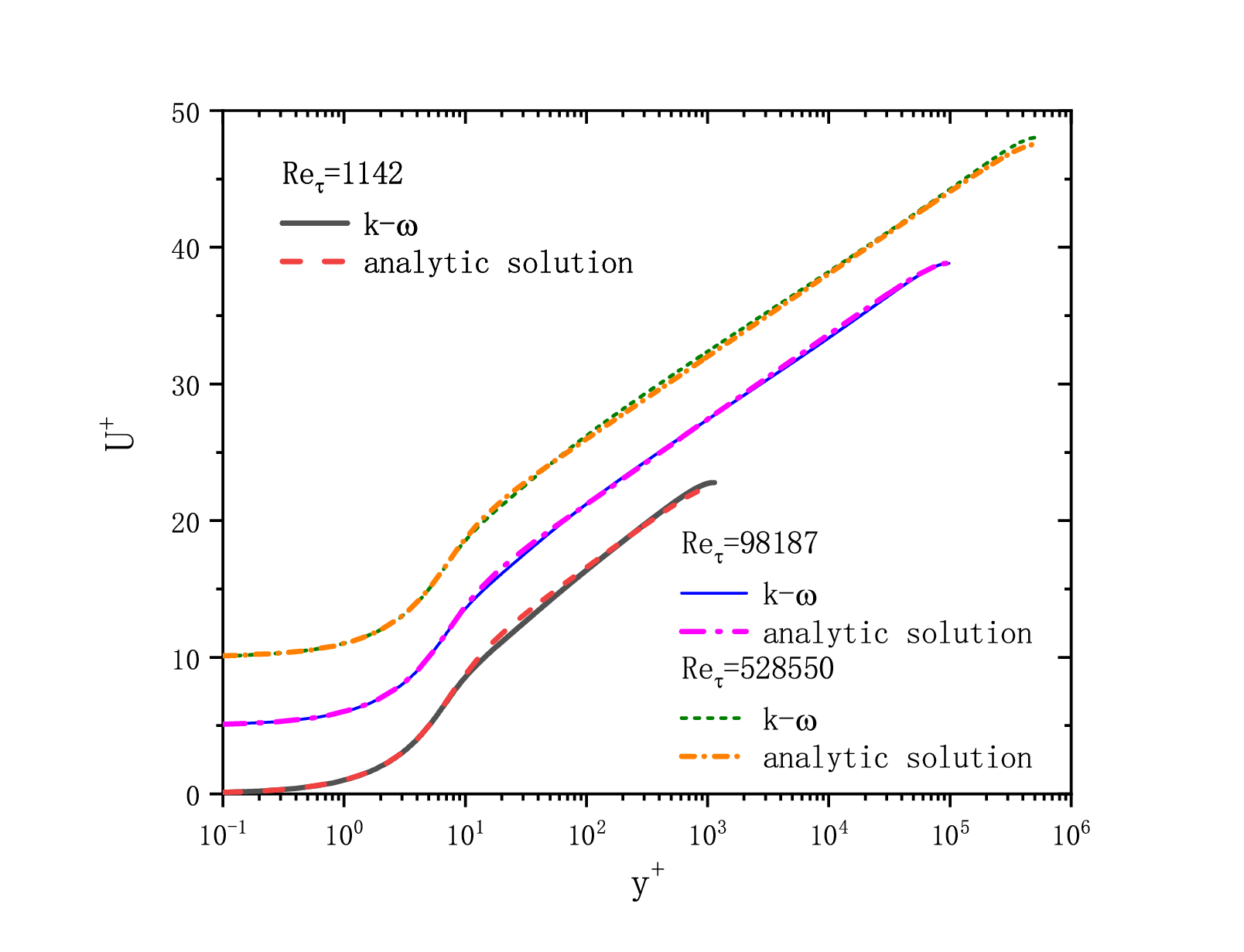}}\quad
\subfigure[]{\includegraphics[width=75mm]{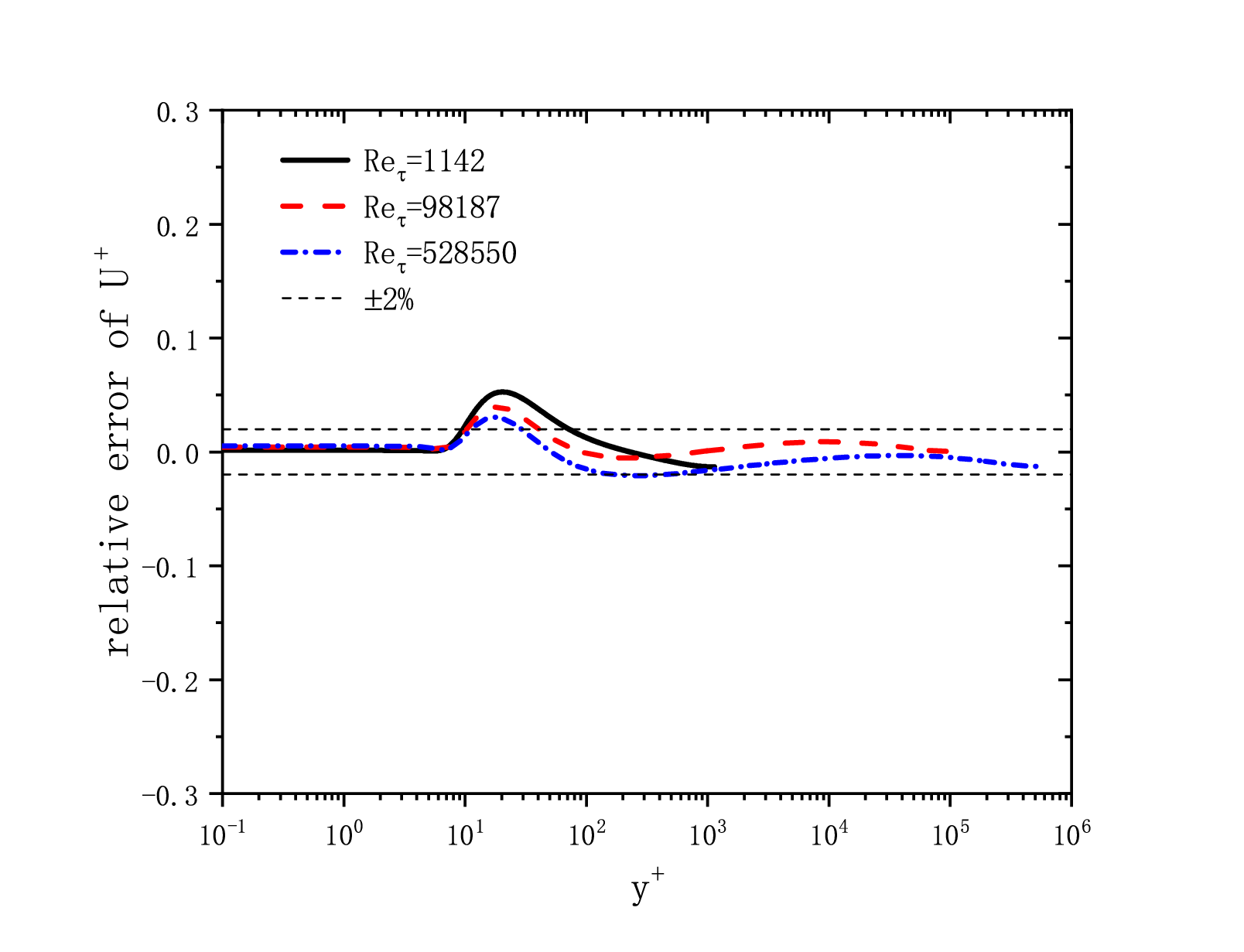}}}
\figcaption{(a) Comparisons between the numerical simulation results of the $k\mbox{-}\omega$ model equations and the predictions of the analytic solutions for $U^+$. (b) The relative error. To display clearly the profiles are vertically shifted for $Re_\tau=98187$ and $528550$ in (a).}\label{fig:U-valid}
\end{figure}

Figure \ref{fig:omega-nut-valid} compares the analytic solutions of the $\omega^+$ and $\nu_t^+$ profiles with the numerical simulation data. The agreement is excellent for both $\omega^+$ and $\nu_t^+$ over the whole pipe radius and over the wide-spread Reynolds number. Fig. \ref{fig:U-valid} shows the comparisons for the streamwise mean velocity profiles. The analytic solutions agree with the numerical results at a considerably high accuracy, as shown further in Fig. \ref{fig:U-valid}(b) by plotting the relative differences with respect to the numerical simulations. Indeed, the relative differences are within $2\%$ over the whole pipe radius except in the buffer layer, where the maximum differences are about $5\%$ at $Re_\tau=1142$, but reduce when $Re_\tau$ is increased.

A final observation is regarding the asymptotic behavior of $\kappa$ and $\kappa_k$. As shown in Table \ref{tab1:parameters}, at $Re_\tau$ over half a million, the measured $\kappa$ and $\kappa_k$ still do not take the values set by the model (0.4 and 1.03), revealing a fact that a rather large $Re_\tau$ is needed to observe the saturation of $\kappa$ and $\kappa_k$. In true pipe flows, similar phenomenon has been observed. For example, at $Re_\tau$ being about half a million\supercite{SuperPipe1998}, $\kappa$ is measured 0.43, which is clearly larger than the conventional recognition coming from experiments at small and moderate $Re_\tau$, approaching but still different from 0.45 predicted by the SED theory\supercite{ChenX-WeiBB}.

\section{Further application to the SED $k\mbox{-}\omega$ model}\label{sec:improved model}

At first, let us summarize the incorrect settings in the multi-layer structure of the $k\mbox{-}\omega$-predicted pipe flow comparing with that of the SED-predicted one, to discuss possible improvements for the $k\mbox{-}\omega$ model. As listed in Table \ref{tab1:parameters}, the incorrect settings include:
\par(a) A power-law exponent of 2, instead of 1.5, for $\ell_{12}^+$ in the viscous sublayer;
\par(b) Power-law exponents of about 3.5 and 2 for $\ell_{12}^+$ and $\ell_k^+$ in the buffer layer, larger than 2 and 1.5 in the SED theory;
\par(c) $y_{sub}^+$ different from $y_{ksub}^+$, and $y_{buf}^+$ different from $y_{kbuf}^+$, all significantly smaller than the corresponding values in the SED theory;
\par(d) $\kappa=0.4$ versus $\kappa=0.45$ in the SED theory;
\par(e) $m=2$ and $r_c=0.4$, in contrast to $m=5$ and $r_c=0.27$ in the SED theory;
\par(f) Incorrect behavior above the buffer layer for $k$.
\\Note that (a) leads to an incorrect scaling of $\nu_t^+ \propto y^{+4}$, instead of $\nu_t^+ \propto y^{+3}$, when approaching the wall, which is known but neglected in Wilcox's analysis\supercite{Wilcox2006} on the $k\mbox{-}\omega$ model equations. Among the above items, (a)-(c) belong to the inner property of boundary layer, and (d)-(e) affect the bulk flow behavior.

Regarding (d)-(e), Chen et al.\supercite{SED-k-omega,ChenX-WeiBB} have proposed the so-called ``SED $k\mbox{-}\omega$" model. They introduced three modifications to the $k\mbox{-}\omega$ model equations:
\par(1) $\alpha_\infty$ and $\sigma$ are changed to $\alpha_\infty=0.57$ and $\sigma=0.32$. Consequently, $\kappa=0.45$ according to Eq. (\ref{eq:kappa}).
\par(2) $\sigma^*$ and $\sigma$ are changed to $\sigma_{SED}^*$ and $\sigma_{SED}$, as: $\sigma_{SED}^*=\sigma^*[1+(\gamma \nu_t^+)^2]$ and $\sigma_{SED}=\sigma[1+(\gamma \nu_t^+)^2]$, where $\gamma=25$. This adjustment increases the turbulent transport in the pipe bulk and remedies the incorrect setting (e) in the $k\mbox{-}\omega$ model equations, as shown below.
\par(3) The dissipations terms in the $k$ and $\omega$ equations are revised to: $-\frac{\beta^*}{\eta^2}k^+\omega^+$ and $-\frac{\beta}{\eta^2}k^+\omega^+$, respectively, where
\begin{align}
\label{eq:sigma_SED} &\eta=c'\left[1+\left(\frac{y^+}{y_B^+}\right)^2\right]^{\gamma_b/2}\left[1+\left(\frac{y^+}{y_M^+}\right)^2\right]^{(\gamma_m-\gamma_b)/2}
\end{align}
with $y_B^+=40$, $y_M^+=\sqrt{Re_\tau/\kappa}$, $\gamma_b\approx0.05$ and $\gamma_m\approx-0.09$ at large Reynolds numbers,
$c'=y_B^{+\gamma_b}c(\beta_0^*/4)^{1/4}$, and $c$ is Reynolds-number-dependent and related to the magnitude of the outer peak of $k$\supercite{SEDPart2,SED-k-omega,ChenX-WeiBB}. These modifications yield a more accurate (above 99\%) description of mean velocity profiles in Princeton super-pipe data for a wide range of Reynolds numbers, improving the Wilcox $k\mbox{-}\omega$ model prediction by up to 10\%. Moreover, they yield an accurate prediction of the entire streamwise mean kinetic energy profiles, where the newly observed outer peak\supercite{SuperPipe2012} is also captured. With a slight change of the wake parameter, the model yields also quite good predictions for turbulent channels and TBLs.

\begin{figure}[h]
\centering \mbox{ \subfigure[]{\includegraphics[width=75mm]{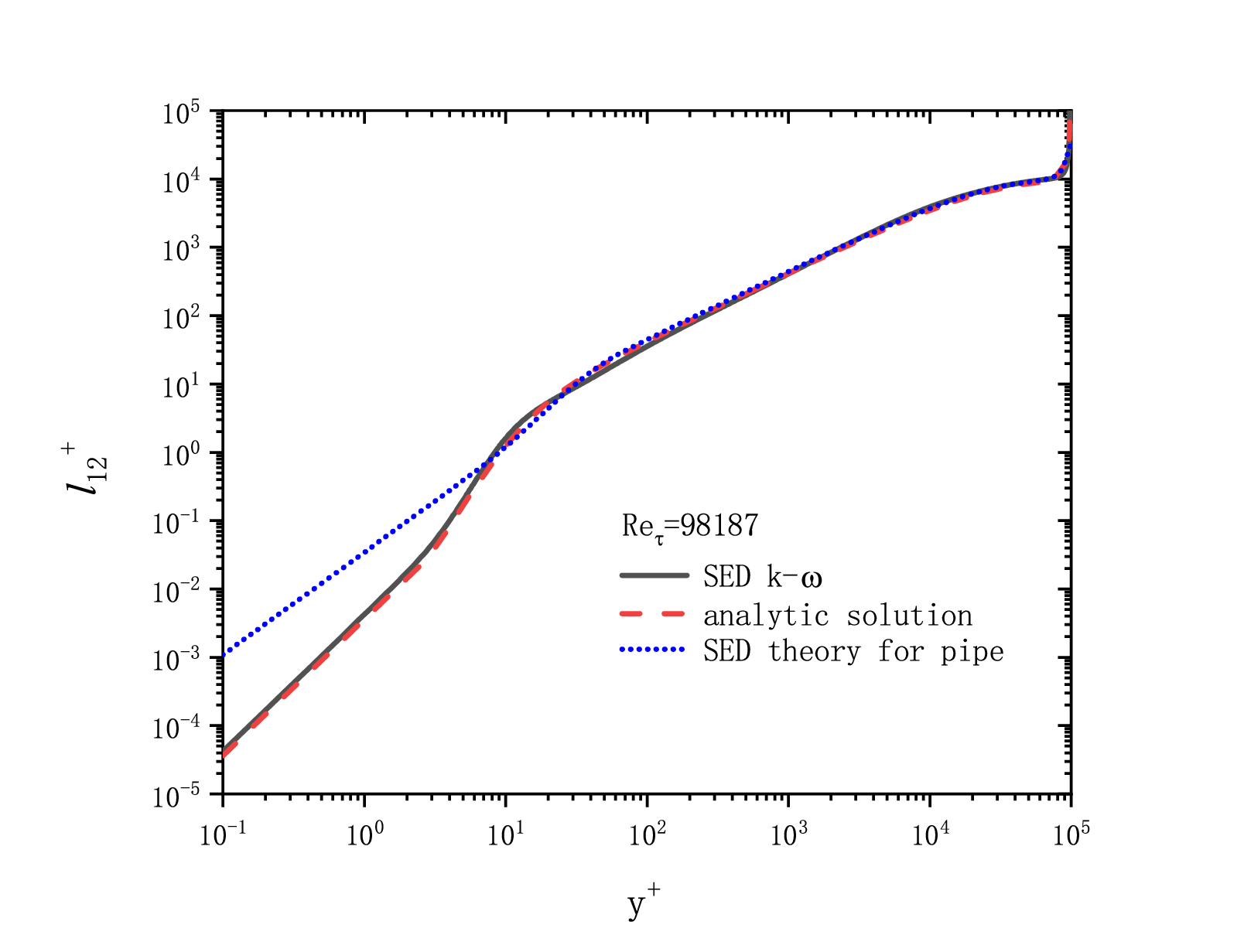}}\quad
\subfigure[]{\includegraphics[width=75mm]{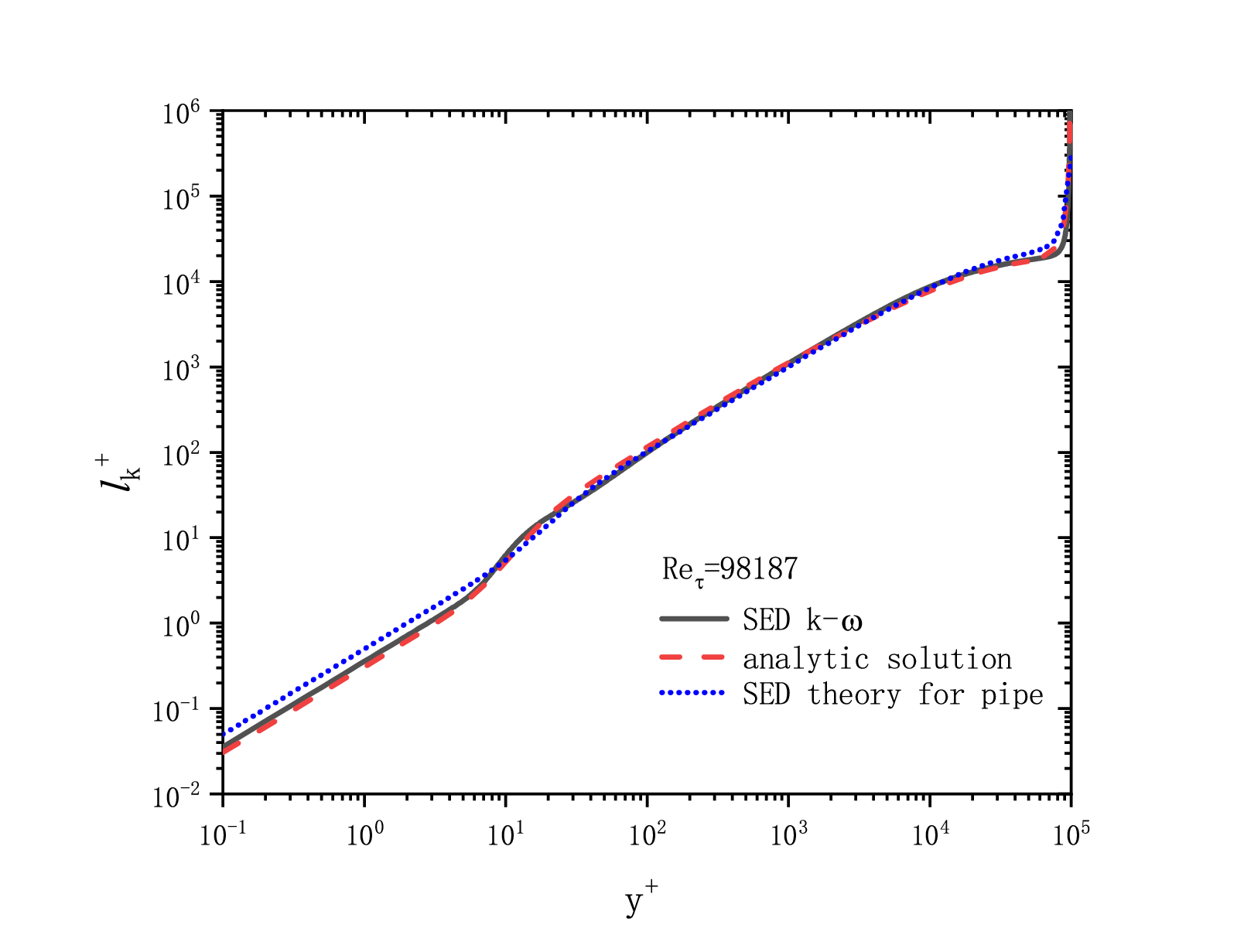}}}
\figcaption{(a) $\ell_{12}^+$ and (b) $\ell_{k}^+$ calculated by the numerical simulation of the SED $k\mbox{-}\omega$ model, compared with the predictions of the analytic solution and the SED theory for pipe.}\label{fig:SEDkw-ell12-ellk-valid}
\end{figure}

In the SED $k\mbox{-}\omega$ model, the inner-layer settings of the Wilcox $k\mbox{-}\omega$ model are preserved because their influences on the mean profiles are weak comparing with the bulk-flow corrections. Here we apply our analytic solution to the SED $k\mbox{-}\omega$ model. In Fig. \ref{fig:SEDkw-ell12-ellk-valid} the numerical results of $\ell_{12}^+$ and $\ell_{k}^+$ calculated by the SED $k\mbox{-}\omega$ model are compared with the current analytic solutions with empirically-estimated multi-layer structure parameters (listed in Table \ref{tab1:parameters}), and with the predictions of the SED theory for pipe\supercite{SEDPart2}. The analytic solutions agree with the numerical simulations quite well for both $\ell_{12}^+$ and $\ell_{k}^+$, showing that the multi-layer structure also is possessed by the SED $k\mbox{-}\omega$ model. Furthermore, whereas $\ell_{k}^+$ of the SED $k\mbox{-}\omega$ model is quite close to that of the SED theory for the whole pipe radius, significant discrepancy occurs in the viscous sublayer of $\ell_{12}^+$, owing to the incorrect scaling law there set by the $k\mbox{-}\omega$ model and preserved by the SED $k\mbox{-}\omega$ model.

\begin{figure}[h]
\centering \mbox{ \subfigure[]{\includegraphics[width=75mm]{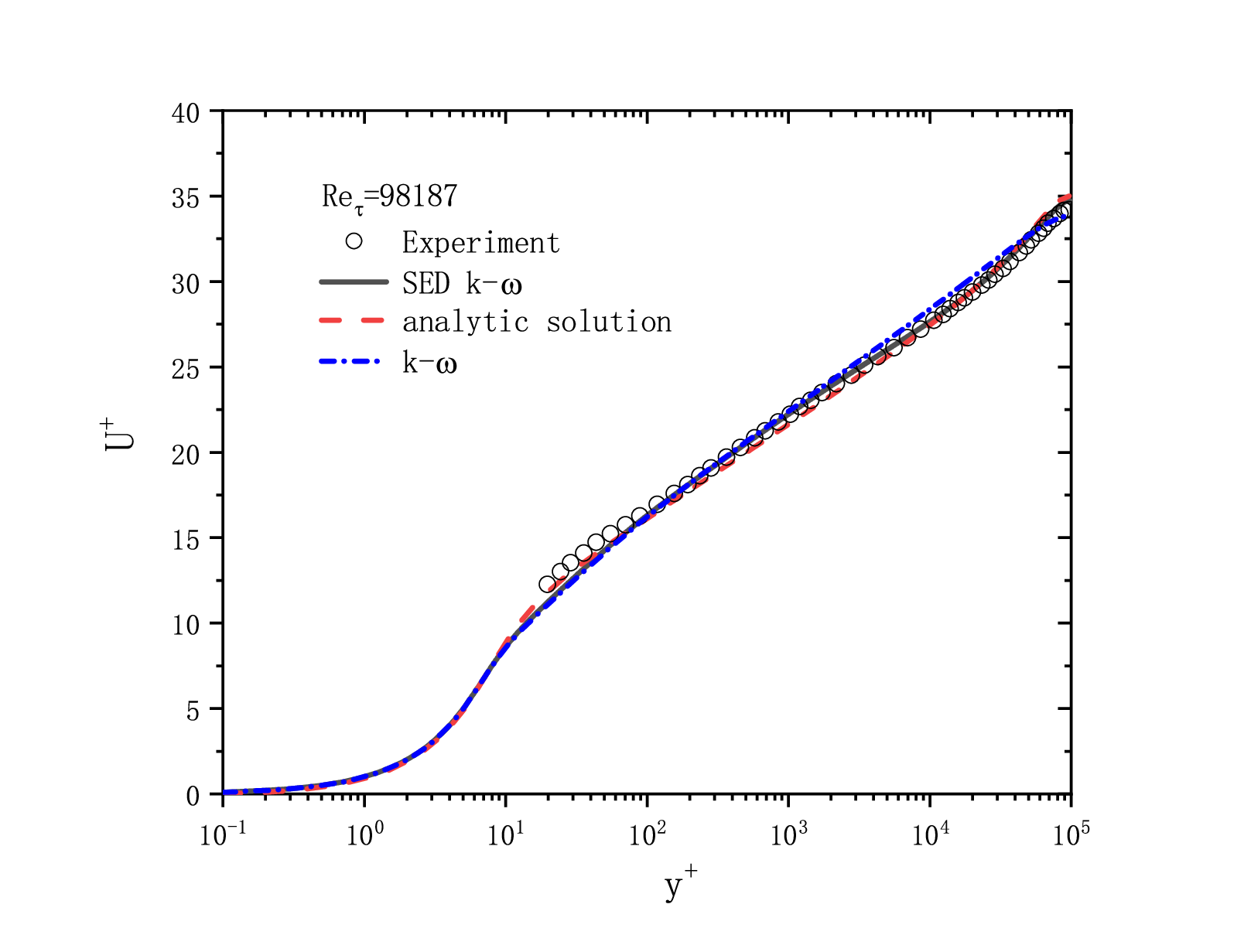}}\quad
\subfigure[]{\includegraphics[width=75mm]{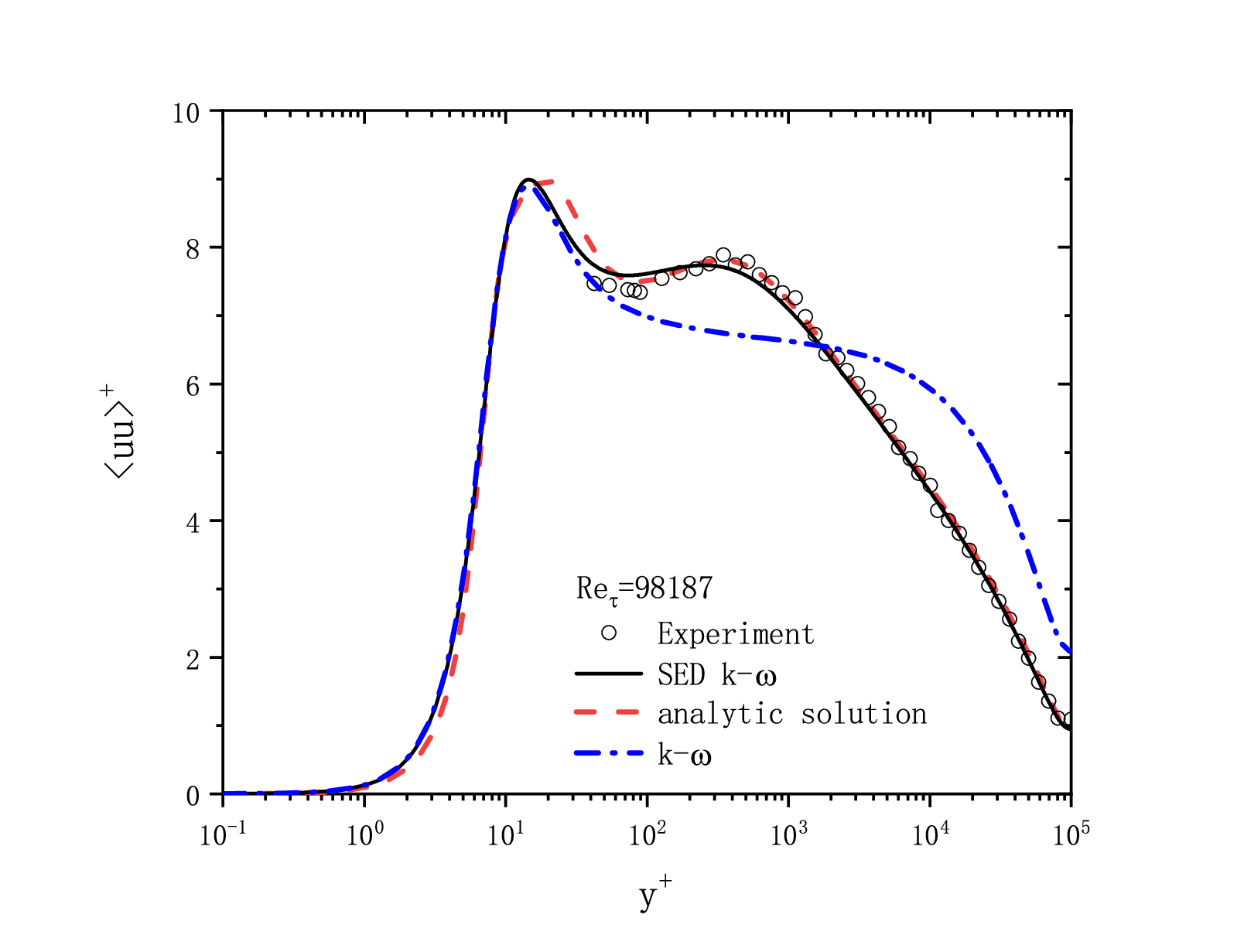}}}
\figcaption{(a) $U^+$ and (b) ${\left \langle uu \right \rangle}^+$ ($=2k^+$) calculated by the numerical simulation of the SED $k\mbox{-}\omega$ model, compared with the predictions of the analytic solution, the results of the $k\mbox{-}\omega$ model, and the experimental data\supercite{SuperPipe2012}.}\label{fig:SEDkw-U-k-valid}
\end{figure}

Figure \ref{fig:SEDkw-U-k-valid}(a) compares the numerical result of $U^+$ calculated by the SED $k\mbox{-}\omega$ model and the prediction of the analytic solution. The two profiles agree with each other quite well, and are rather close to the experimental results\supercite{SuperPipe2012}, superior than those of the $k\mbox{-}\omega$ model are. Figure \ref{fig:SEDkw-U-k-valid}(b) further compares the ${\left \langle uu \right \rangle}^+$ ($=2k^+$) profiles, with together the $k\mbox{-}\omega$ model results and the experimental data. The analytic solution differs quite little from the numerical result of the SED $k\mbox{-}\omega$ model, and the SED $k\mbox{-}\omega$ model accurately reproduces the experimental profile of ${\left \langle uu \right \rangle}^+$ , especially for the outer peak that is not captured by the $k\mbox{-}\omega$ model. Note that in predicting ${\left \langle uu \right \rangle}^+$ via Eqs. (\ref{eq:ellkformula}) and (\ref{eq:Splus}), the right hand side of (\ref{eq:ellkformula}) is multiplied by $\eta$, which characterizes an abnormal scaling law of $k^+$ in a mesolayer of wall turbulence\supercite{SEDPart2,SED-k-omega,ChenX-WeiBB}.

The multi-layer structure parameters of the SED $k\mbox{-}\omega$ model equations are listed in Table \ref{tab1:parameters}. One finds that $\kappa$ and $\kappa_k$ possess also a finite Reynolds number effect. The viscous sublayer thicknesses of $\ell_{12}^+$ and $\ell_{k}^+$ are the same for the SED $k\mbox{-}\omega$ model and the $k\mbox{-}\omega$ model, and the buffer layer thicknesses are only slightly different, revealing a similar inner-layer setting for the two models. On the other hand, in the SED $k\mbox{-}\omega$ model, the center core size and the defect law are agreeing with the prediction of the SED theory for pipe, showing that the incorrect setting in the pipe bulk by the $k\mbox{-}\omega$ model has been remedied.

\section{Discussions and conclusions}\label{sec:conclusion}

This work sets out an almost impossible goal to formulate an analytic solution to the highly nonlinear $k$ and $\omega$ equations of the Wilcox $k\mbox{-}\omega$ model; but remarkably, the multi-layer analytic form of two lengths, Eq. (\ref{eq:ell12formula}) and Eq. (\ref{eq:ellkformula}), are proved to be extremely accurate in predicting the numerical solutions of the $k\mbox{-}\omega$ model equations at three different $Re_\tau$ covering several orders of magnitudes and for all relevant quantities such as Reynolds stress profile, $k$ profile, $\omega$ profile, eddy viscosity profile, and most importantly, the mean velocity profile, as shown in Fig. (\ref{fig:W-uu-valid}), Fig. (\ref{fig:omega-nut-valid}) and Fig. (\ref{fig:U-valid}). Note that in the predictions, the multi-layer parameters are completely invariant with $Re_\tau$, except for $\kappa$ and $\kappa_k$, which seem to possess a finite $Re$ effect, as listed in Table \ref{tab1:parameters}. This finite $Re$ effect explains why, for years, the measured $\kappa$ for data in Princeton Superpipe is around 0.42-0.43, which is fully consistent with an asymptotic $\kappa=0.45$ predicted by the SED theory.

What is the significance of such a set of analytic descriptions of the $k\mbox{-}\omega$ model? First, it confirms that there exists a similarity solution in $Re$ in the turbulent regime as well as in the laminar regime. For laminar pipe, the similarity solution is the simple parabolic mean velocity profile; then, current work shows that in the turbulent regime, it is the multi-layer profile of the two lengths which are invariant with $Re$ (e.q. Eq. (\ref{eq:ell12formula}) and Eq. (\ref{eq:ellkformula})). Similarity solution for flat-plate TBL may be more important for applications, for which the laminar regime is described by the Blasius solution; we have shown\supercite{SEDPart1,SEDPart2} that the multi-layer form is also valid, at least at high $Re$. So, we assert that the long-sought similarity solution of turbulent wall flow is in fact the multi-layer solution given by the SED theory, which is currently shown to also be the mathematical structure of the $k\mbox{-}\omega$ model. Engineers capture this structure by genuinely tuning the parameter values and (three) transition functions, which has remained implicit during the past several decades, but now made explicit. For all, this revelation would demystify the success of the $k\mbox{-}\omega$ model.

Secondly, the success of the current description of the $k\mbox{-}\omega$ solutions demonstrate more clearly the validity of the generalized dilation symmetry-breaking ansatz, which should stand as the self-organization principle for near-wall eddies. The reasons for this assertion are two-fold. On one hand, it may not be surprising to anyone, who deeply understands the essence of physics being just space and time, to see the importance of eddy lengths, but it is remarkable to see that once they are chosen as similarity variables, the simple ansatz (Eq. (\ref{eq:ell12formula}) and Eq. (\ref{eq:ellkformula})) directly gives rise to solutions for turbulent wall flows, which still surprises many people. Now, accurately using eddy length functions to solve the $k\mbox{-}\omega$ model equations, which have $k$ and $\omega$ as the variables, demonstrates unarguably the role of the lengths which quantify dominant eddy sizes. So, the current analysis touches upon the basic physics of wall turbulence: wall-related symmetry constraint on eddy sizes is the governing principle of wall turbulence. On the other hand, the elegant functional form in Eq. (\ref{eq:ell12formula}) and Eq. (\ref{eq:ellkformula}) for mathematically describing the physical multi-layer structure of wall turbulence indicates an ultimate truth: the simplicity is the rule. These two features are not specific to pipe or any particular wall flows, but general to all wall flows, as we have shown in a serious of explorations, from flat plate to airfoil and cone flows\supercite{SEDSLSC,SEDSLAMS,SEDSLAIA}, from subsonic to hypersonic flow\supercite{SEDSLJFM}, and to rough wall flow\supercite{SEDNJP}, up to atmospheric surface layers\supercite{SEDAtmos}. Therefore, the multi-layer form reflects a self-organization principle which governs the variation of the similarity variables after the (dilation-) symmetry-breaking. We think that this recognition may yield surprising outcomes such as what is recently accomplished in high temperature superconductivity studies\supercite{LRNJP,LRCP}.

In the previous development of the SED theory, the most important multi-layer parameters, $y^+_{sub}$ and $y^+_{buf}$ (also $y^+_{ksub}$ and $y^+_{kbuf}$) were obtained, purely empirically, by fitting experimental and numerical data. In the current work, we have derived four relations, (e.g. Eq. (\ref{eq:ybuf-est}), Eq. (\ref{eq:kbuf-approx}), Eq. (\ref{eq:sublayer-constraint1}), and Eq. (\ref{eq:sublayer-constraint0})), which enable us to estimate the four layer thicknesses. Although the estimated values are slightly away from empirically-determined values, which can be further improved by iterations to include higher-order correction (in the future), the current work is making a breakthrough in calculating these critical parameters, similar to our success in deriving the Monin-Obukhov similarity function\supercite{SEDAtmos}. It is intriguing to pursue this path in the N-S TBLs, since we have the expression valid throughout the entire domain. Although the local analysis is performed by inserting local approximation, the constraint relations come from the global constraint by the wall, which is here named as globally-directed local analysis. This method may be useful in solving other highly nonlinear problems with a good guess on the global structure.



\renewcommand{\theequation}{A\arabic{equation}}
\setcounter{equation}{0}

%

\end{CJK*}

\begin{thebibliography}{00}   
\footnotesize
\bibitem{CFD2030}SLOTNICK, J., KHODADOUST, A., ALONSO, J., et al. CFD vision 2030 study: A path to revolutionary computational aerosciences. \textit{NASA CR}, 218178 (2014)
\bibitem{Wilcox2006}WILCOX, D. C. \emph{Turbulence modeling for CFD}, 3rd ed., DCW Industries, California, 124--128 (2006)
\bibitem{Spalart2006}SPALART, P. Turbulence. Are we getting smarter? In: \emph{Fluid Dynamics Award Lecture, 36th Fluid Dynamics Conference and Exhibit}, San Francisco, CA (2006)
\bibitem{Kolmogorov}KOLMOGOROV, A. N. The equation of turbulent motion in an incompressible viscous fluid. \textit{Izv Akad Nauk SSSR}, \textbf{VI}, 56--58 (1942)
\bibitem{Saffman}SAFFMAN P. G. A model for inhomogeneous turbulent flow. \textit{Proc. R. Soc. Lond.}, \textbf{A317}, 417--433 (1970)
\bibitem{Spalding}LAUNDER, B. E. and SPALDING, D. B. \emph{Mathematical models of turbulence}, Academic Press, Landon (1972)
\bibitem{MenterSST}MENTER, F. R. Two-equation eddy-viscosity turbulence models for engineering applications. \textit{AIAA J.}, \textbf{32}(8), 1598--1605 (1994)
\bibitem{SEDPart1}SHE, Z. S., CHEN, X., and HUSSAIN, F. Quantifying wall turbulence via a symmetry approach: A Lie group theory. \textit{J. Fluid Mech.}, \textbf{827}, 322--356 (2017)
\bibitem{SEDPart2}CHEN, X., HUSSAIN, F., and SHE, Z. S. Quantifying wall turbulence via a symmetry approach. Part 2. Reynolds stresses. \textit{J. Fluid Mech.}, \textbf{850}, 401--438 (2018)
\bibitem{SEDSLSC}XIAO, M. J. and SHE Z. S. Symmetry-based description of laminar-turbulent transition. \textit{Sci. China: Phys. Mech. Astron.}, \textbf{62}(9), 994711 (2019)
\bibitem{LiuF2021}LIU, F., FANG, L., and FANG, J. Non-equilibrium turbulent phenomena in transitional flat plate boundary-layer flows. \textit{Appl. Math. Mech. - English Edition}, \textbf{42}(4), 567-582 (2021)
\bibitem{SED-k-omega}CHEN, X., HUSSAIN, F., and SHE, Z. S. Predictions of canonical wall-bounded turbulent flows via a modified $k�C\omega$ equation. \textit{J. Turbul.}, \textbf{18}(1), 1--35 (2017)
\bibitem{DongM2017}YE, M. S. and DONG, M. Near-wall behaviors of oblique-shock-wave/turbulent-boundary-layer interactions. \textit{Appl. Math. Mech. - English Edition}, \textbf{38}(10), 1357-1376 (2017)
\bibitem{WuXH-DNS}Wu, X. H. and MOIN, P. A direct numerical simulation study on the mean velocity characteristics in turbulent pipe flow. \textit{J. Fluid Mech.}, \textbf{608}, 81--112 (2008)
\bibitem{SuperPipe1998}ZAGAROLA, M. V. and SMITS, A. J. Mean-flow scaling of turbulent pipe flow. \textit{J. Fluid Mech.}, \textbf{372}, 33-79 (1998)
\bibitem{SuperPipe2012}HULTMARK, M., VALLIKIKV, M., BAILEY, S. C. C., et al. Turbulent pipe flow at extreme Reynolds numbers. \textit{Phys. Rev. Lett.}, \textbf{108}, 094502 (2012)
\bibitem{SED-kappa}WU, Y., CHEN, X., SHE, Z. S., and HUSSAIN, F. On the Karman constant in turbulent channel flow. \textit{Phys. Scripta.}, 2013:014009 (2013)
\bibitem{ChenX-WeiBB}CHEN, X., WEI, B. B., HUSSAIN, F., and SHE Z. S. Anomalous dissipation and kinetic-energy distribution in pipes at very high Reynolds numbers. \textit{Phys. Rev. E}, \textbf{93}, 011102(R) (2015)
\bibitem{SEDSLAMS}XIAO, M. J. and SHE Z. S. Precise drag prediction of airfoil flows by a new algebraic model. \textit{Acta Mech. Sinica}, \textbf{36}(1), 35--43 (2020)
\bibitem{SEDSLAIA}BI, W. T., WEI, Z., ZHENG, K. X., and SHE Z. S. A symmetry-based length model for characterizing the hypersonic boundary layer transition on a slender cone at moderate incidence. \textit{Adv. in Aerodyn.}, \textbf{4}:26, 1--23 (2022)
\bibitem{SEDSLJFM}SHE Z. S., ZOU, H. Y., XIAO, M. J., et al. Prediction of compressible turbulent boundary layer via a symmetry-based length model. \textit{J. Fluid Mech.}, \textbf{857}, 449--468 (2018)
\bibitem{SEDNJP}SHE Z. S., WU, Y., CHEN, X., and HUSSAIN, F. A multi-state description of roughness effects in turbulent pipe flow. \textit{New J. Phys.}, \textbf{14}, 093054 (2012)
\bibitem{SEDAtmos}JI, Y. and SHE Z. S. Analytic derivation of Monin-Obukhov similarity function for open atmospheric surface layer. \textit{Sci. China: Phys. Mech. Astron.}, \textbf{64}(3), 34711 (2021)
\bibitem{LRNJP}LI, R. and SHE, Z. S. Emergent mesoscopic quantum vortex and Planckian dissipation in the strange metal phase. \textit{New J. Phys.}, \textbf{23}(4), 043050 (2021)
\bibitem{LRCP}LI, R. and SHE, Z. S. Unified energy law for fluctuating density wave orders in cuprate pseudogap phase. \textit{Comm. Phys.}, \textbf{5}(1), 13 (2022)
\end{thebibliography}
\end{document}